\documentclass[useAMS,usenatbib]{mnras}
\usepackage{graphicx}
\usepackage[normalem]{ulem}
\usepackage{threeparttable}

\newcommand{\zsol}{Z$_{\odot}$}
\newcommand{\msol}{M$_{\odot}$}

\newcommand{\hii}{H\,{\small{\sc ii}}}
\newcommand{\ha}{H$\alpha$}
\newcommand{\hb}{H$\beta$}
\newcommand{\hg}{H$\gamma$}
\newcommand{\Av}{$A_{\rm V}$}

\newcommand{\heiiwr}{He\,{\sc ii}\,$\lambda4686$}

\newcommand{\niiiwr}{N\,{\sc iii}\,$\lambda\lambda4634/41$}
\newcommand{\nvwr}{N\,{\sc v}\,$\lambda\lambda4603/20$}
\newcommand{\heineb}{He\,{\sc i}\,$\lambda4713$}

\newcommand{\civwrr}{C\,{\sc iv}\,$\lambda\lambda5801/12$}

\newcommand{\ciiiwrb}{C\,{\sc iii}\,$\lambda\lambda4647/66$}
\newcommand{\civwrb}{C\,{\sc iv}$\lambda4658$}

\newcommand{\feiii}{[Fe\,{\sc iii}]\,$\lambda4658$}

\newcommand{\ergs}{erg\,s$^{-1}$}
\newcommand{\ergcms}{erg\,cm$^{-2}$\,s$^{-1}$}
\newcommand{\ergcmsang}{erg\,cm$^{-2}$\,s$^{-1}$\,\AA$^{-1}$}
\newcommand{\ergcmsarc}{erg\,cm$^{-2}$\,s$^{-1}$\,arcsec$^{-2}$}

\newcommand{\hei}{He\,{\sc i}}
\newcommand{\heii}{He\,{\sc ii}}

\newcommand{\hst}{{\it HST}}

\title[He\,II emission in NGC\,1569]
{MEGARA-IFU detection of extended \heiiwr\ nebular emission in the central region of NGC\,1569 and its 
ionization budget}
\author[Y.\,D.\,Mayya et al.]{Y.\,D.\,Mayya$^{1}$\thanks{Email: ydm@inaoep.mx}, E. Carrasco$^{1}$, 
V.M.A. G\'omez-Gonz\'alez$^{2}$, J. Zaragoza-Cardiel$^{1,3}$,
\newauthor 
A. Gil de Paz$^{4}$, P. A. Ovando$^{1}$, M. S\'anchez-Cruces$^{5}$, L. Lomel\'{\i}-N\'u\~nez$^{1}$, 
\newauthor 
L. Rodr\'{\i}guez-Merino$^{1}$, D. Rosa-Gonz\'alez$^{1}$, S. Silich$^{1}$, G. Tenorio-Tagle$^{1}$, 
\newauthor 
G. Bruzual$^{2}$, S. Charlot$^{6}$, R. Terlevich$^{1,7}$, E. Terlevich$^{1}$, O. Vega$^{1}$, J. Gallego$^{4}$, 
\newauthor 
J. Iglesias-P\'aramo$^{8,9}$, A. Castillo-Morales$^{4}$, M.L. Garc\'{\i}a-Vargas$^{10}$, 
\newauthor 
P. G\'omez-Alvarez$^{10}$, S. Pascual$^{4}$ and A. P\'erez-Calpena$^{10}$
\\
$^{1}$Instituto Nacional de Astrof{\'\i}sica, \'Optica y Electr\'onica, Luis Enrique Erro 1, Tonantzintla 72840, Puebla, Mexico\\
$^{2}$Instituto de Radioastronom\'{i}a y Astrof\'{i}sica, UNAM Campus Morelia, Apartado postal 3-72, 58090 Morelia, Michoac\'an, Mexico\\
$^{3}$ Consejo Nacional de Ciencia y Tecnolog\'ia, Av. Insurgentes Sur 1582, 03940,  Mexico City, Mexico\\
$^{4}$Dpto. de F\'{i}sica de la Tierra y Astrof\'{i}sica, and Instituto de F\'{i}sica de Part\'{i}culas y del Cosmos (IPARCOS), Fac. CC. F\'{i}sicas, \\
Universidad Complutense de Madrid, Plaza de las Ciencias, 1, E-28040 Madrid, Spain \\
$^{5}$Aix Marseille Univ., CNRS, CNES, LAM, Marseille, France\\ 
$^{6}$Sorbonne Universit\'e, CNRS, UMR7095, Institut d'Astrophysique de Paris, F-75014, Paris, France \\
$^{7}$Institute of Astronomy, University of Cambrige, Cambridge CB3 0HA, UK \\
$^{8}$Instituto de Astrof\'{i}sica de Andaluc\'{i}a, IAA-CSIC, Glorieta de la Astronom\'{i}a s/n, E-18008 Granada, Spain \\
$^{9}$Estaci\'on Experimental de Zonas \'Aridas, CSIC, Carretera de Sacramento s/n, E-04120 Almer\'{i}a, Spain\\ 
$^{10}$FRACTAL S.L.N.E., Calle Tulip\'an 2, portal 13, 1A, E-28231 Las Rozas de Madrid, Spain \\
}

\begin{document}
\maketitle

\begin{abstract}
We here report the detection of extended \heiiwr\ nebular emission in the 
central region of NGC\,1569 using the integral field spectrograph MEGARA at 
the 10.4-m {\it Gran Telescopio Canarias}. The observations cover 
a Field of View (FoV) of 12.5~arcsec$\times$11.3~arcsec at seeing-limited 
spatial resolution of $\sim$15~pc and at a spectral resolution of R=6000 in 
the wavelength range 4330--5200\,\AA. The emission extends over a semi-circular 
arc of $\sim$40~pc width and $\sim$150~pc diameter around the super star 
cluster A (SSC-A). The \Av\ derived using Balmer decrement varies from the 
Galactic value of 1.6~mag to a maximum of $\sim$4.5~mag, with a mean value 
of 2.65$\pm$0.60~mag. We infer 124$\pm$11 Wolf-Rayet (WR) stars in SSC-A using 
the \heiiwr\ broad feature and \Av=2.3~mag. The He$^+$ ionizing photon rate 
from these WR stars is sufficient to explain the luminosity of the 
\heii\ nebula. The observationally-determined total He$^+$ and H$^0$ ionizing 
photon rates, their ratio, and the observed number of WR stars in SSC-A are 
all consistent with the predictions of simple stellar population models at an age 
of 4.0$\pm$0.5~Myr, and mass of (5.5$\pm0.5$)$\times10^5$~\msol. Our 
observations reinforce the absence of WR stars in SSC-B, the second most massive 
cluster in the FoV. None of the other locations in our FoV where \heiiwr\ emission 
has been reported from narrow-band imaging observations contain WR stars.
\end{abstract}


\section{Introduction}

Availability of large telescopes equipped with integral field spectrographs has
enabled the creation of maps of \heiiwr\ nebular emission in nearby star-forming 
galaxies \citep[e.g.][]{Kehrig2015, Kehrig2018}. Given that only photons 
shortward of 228~\AA\ (54 eV) can doubly ionize helium, the \heiiwr\ line 
provides a ground-based tool to trace the hard part of the ultraviolet (UV)
spectrum. High mass stars, especially during their short-duration Wolf-Rayet 
(WR) phase, are the most common sources that emit these hard UV photons \citep{Schaerer1996}. 
The UV flux emerging from a starburst region depends both on the hardness of
the UV spectrum of WR stars as well as on the number of WR stars present.
Because of the presence of an expanding atmosphere, models incorporating 
specialized line-blanketing treatment are required in order to quantitatively 
predict the flux of hard UV photons emitted by the WR stars. The emergent spectra 
in models incorporating such treatment \citep[e.g.][]{Hillier1998, Grafener2002}
are found to be softer as compared to earlier models that did not take into 
account opacities from metals \citep{Schmutz1992}.
On the other hand, the number of WR stars and the duration of the WR phase in 
a starburst region depend on metallicity \citep[e.g.][]{Maeder1989, Geneva1992, Padova1994, Chen2015}, 
stellar rotation \citep{Meynet2005} and the stellar multiplicity \citep{Eldridge2017}. 
The calculated rate of He$^+$ ionizing photons at different metallicities 
depends critically on the mass-loss recipes used during massive star evolution. 
The general trend from the different set of currently available codes is a decrease in 
the expected luminosity of the \heiiwr\ line with decreasing metallicity.

The spectra from Sloan Digital Sky Survey (SDSS) have enabled the detection of 
the relatively faint \heiiwr\ line in large samples of star-forming galaxies 
\citep[e.g.][]{Shirazi2012}. These studies find that the observed \heiiwr/\hb\ 
intensity ratio does not drop at low metallicities. In fact, recent studies
find the ratio to be increasing as the metallicity decreases \citep{Schaerer2019}.
Furthermore, these low-metallicity \heiiwr-emitting galaxies often show only a 
weak or no evidence of the presence of WR stars \citep{Shirazi2012}. 
Thus, questions have been raised on the WR stars as the sole source of ionization 
of He$^+$ \citep{Plat2019}.
Alternative mechanisms such as hard radiation from high-mass stars in binaries
\citep{Eldridge2017},
shocks from supernova remnants \citep{Garnett1991, Dopita1996} and high-mass X-ray 
binaries \citep[HMXB;][]{Schaerer2019, Kojima2020} are often invoked. 
Nearby low-metallicity systems offer an opportunity to address the He$^+$ ionization
problem by enabling study of individual star-forming knots.
In a detailed study 
of the metal-poor (Z=3--4\% \zsol) galaxy SBS\,0335~$-$~052E using MUSE,
\citet{Kehrig2018} discard WR stars as the source of ionization 
and instead propose rotating metal-free stars or a binary population with 
$Z=10^{-5}$ and an extremely top-heavy initial mass function (IMF) as the only 
plausible way of getting around the problem of the ionization budget.
In a recent study \citet{Schaerer2019} find that the observed \heiiwr\ 
intensity in metal-poor star-forming galaxies can be naturally reproduced if 
the bulk of the He$^+$ ionizing photons is emitted by the HMXB, whose number 
is found to increase with decreasing metallicity. X-ray binaries in a cluster 
appear only after the death of the most massive stars, and hence this scenario cannot 
explain the He$^+$ ionization in young systems (\hb\ equivalent widths (EWs) 
$\geq$200~\AA), as illustrated by \citet{Plat2019}.

A detailed analysis of the \heii\ ionization budget problem has been carried out 
only in a handful of extreme metal-poor galaxies \citep[e.g.][]{Kehrig2011,Kehrig2018}.
Lack of observational data of individual massive stars at these low metallicities 
makes the predictions of population synthesis calculations heavily dependent on 
the theoretical assumptions. On the other hand, model calculations have been 
better calibrated at the LMC and solar metallicities. However, its
detection requires sensitive observations, specially 
aimed at detecting faint emission lines. The newly available spectrograph MEGARA 
at the 10.4-m Gran Telescopio Canarias (GTC), equipped with Integral Field Unit 
(IFU) at spectral resolution $\sim$6000 has the capability of detecting 
and mapping the faint emission lines from extragalactic nebulae \citep{GildePaz2020}. 
In order to exploit this capability, we carried out MEGARA observations of NGC\,1569, a 
dwarf galaxy with a gas-phase oxygen abundance close to that of the LMC 
\citep[$12+\log({\rm O/H})$=8.19;][]{Kobulnicky1997}.
We mapped its central region, which is known to have extended \ha\ emission
\citep{Hodge1974, Waller1991, Hunter1993}.
Throughout this study, we use a distance of 3.1~Mpc, measured  
using Hubble Space Telescope ({\it HST}) observations of the tip of the red 
giant branch \citep{Grocholski2012}. This distance is $\sim$10\% smaller than
that obtained by \citet{Grocholski2008} using an earlier analysis of the same dataset, but
is still considerably higher than the 2.2~Mpc distance 
\citep{Israel1988} that was being routinely used in studies prior to the \citet{Grocholski2008} work.

NGC\,1569 is among the nearest galaxies that harbours young superstellar clusters 
(SSCs) that are as massive as the Galactic globular clusters. Its most massive SSCs, 
called A and B, are estimated to have dynamical masses of 
$4.1\times10^5$~\msol\ \citep{Ho1996} and $6.2\times10^5$ \msol \citep{Larsen2008}, 
respectively \citep[after rescaling the masses to the currently used distance of 3.1~Mpc
for this galaxy from][]{Grocholski2012}. These two SSCs are at the high-mass end of a 
population of around 50 SSCs, whose ages range from a few million to several 
hundred million years \citep{Hunter2000}. The \ha\ morphology of the galaxy is 
dominated by \hii\ regions \citep{Waller1991}, large-scale shells and 
superbubbles \citep*{Hunter1993, Westmoquette2008}, with the brightest \hii\ 
region in this galaxy associated to the star forming complex surrounding 
cluster 10 of \citet{Hunter2000} that lies 105~pc (7~arcsec) to the east of 
SSC-A, and is outside our FoV. Extended X-ray emission is detected in the galaxy, 
most of which coincides with the \ha\ bubbles \citep{Martin2002, Monica2015}.
Giant molecular clouds have been detected, but none of them are associated to 
SSCs A and B \citep{Taylor1999}. The largest CO cloud complex in the galaxy 
lies to the east of cluster 10 outside our FoV. 

Long-slit spectroscopic observations have established the presence of 
WR stars in SSC-A, whereas no WR stars have been detected in SSC-B \citep{Rosa1997}. 
Using narrow-band \hst\ imaging observations with the F469N filter, 
\citet{Buckalew2000} have inferred the presence of WR features from 5 SSCs, 
including SSC-A, and 7 individual stars, and nebular emission from three 
additional point sources. \citet{Kobulnicky1997} reported faint \heii\ nebular 
emission at some of the locations along the longslits they had used. However,
no extended \heiiwr\ nebular emission has been yet detected in this galaxy. In 
this paper, we present our results obtained from spectral mapping of the central
zone that includes both the SSCs A and B.

In Section~2, we describe the observations and the data reduction. Techniques for
separating the \heiiwr\ emission lines of stellar origin from nebular origin 
are also detailed in this section. In Section~3, we present the maps in these two
components and compare them with the maps available at the \hst\ resolution. 
Then, Section~4 deals with the calculation of the number of WR stars and the ionizing 
photon rate of doubly ionized helium. The ionization budget is discussed in Section~5, 
and our conclusions are given in Section~6. Measured data in individual fibre spectrum
are presented in an appendix.

\section{Observations, reduction and data analysis}

\subsection{Observations}

Spectroscopic observations of the central part of NGC\,1569 (RA=04:30:48.5, 
DEC=+64:50:55.5) were carried out using the recently available MEGARA 
instrument at the 10.4-m GTC. MEGARA provides  multi-object and integral 
field spectroscopy at low, medium and high spectral resolutions 
$R_{\rm FWHM}\equiv\lambda/\Delta\lambda\sim$6000, 12000 and 20000, respectively, 
in the visible wavelength interval from 3650 to 9700~\AA, through 18 spectral 
configurations ($\Delta\lambda$=Full Width at Half Maximum (FWHM) of a line).
The IFU, also named Large Compact Bundle (LCB), covers an area on the sky of 
12.5~arcsec$\times$11.3~arcsec with 567 fibres for a spaxel size of 
0.62~arcsec. To perform simultaneous sky subtraction, the IFU fibres are 
supplemented by another 56 fibres distributed in 8 mini-bundles of 7 fibres, 
located at the edge of the field at distances from 1.7~arcmin to 2.5~arcmin 
from the centre of the LCB. The complete set of 623 fibres is mounted in the 
LCB spectrograph pseudo-slit. Additionally, a set of robotic  positioners host 
92 mini-bundles of 7 fibres each, also for a spaxel size of 0.62~arcsec, 
allowing  observations in a field of view (FoV) of 3.5~arcmin$\times$3.5~arcmin
in the multi-object spectrograph (MOS) mode. These 644 fibres are arranged in a 
different pseudo-slit interchangeable with the LCB pseudo-slit. For a complete 
description of MEGARA performance at GTC and the first scientific results 
obtained during the instrument commissioning, see \citet{2018SPIE10702E..16C} 
and \citet{2018SPIE10702E..17G, GildePaz2020}. 

The observations reported here were carried out as part of MEGARA guaranteed 
time on 2019 February 9$^\mathrm{th}$ in queue mode under dark sky and 
photometric conditions. The data were obtained using the IFU mode in 
combination with the LR-B volume phase holographic grating, centred at 4800~\AA, 
covering the spectral range from $\sim$4330 to 5200~\AA\ for a resolution of 
FWHM=0.78\,\AA\ with a reciprocal dispersion of 0.21\,\AA/pixel. Three 1200~s 
exposures were taken at airmasses between 1.25--1.30, and seeing 
$\sim$0.9~arcsec. For flux calibration the spectroscopic standard star HR5501 was 
observed in the same setup as for NGC\,1569. Bias, flat field and arc lamp images 
were also obtained as part of the data package.

\subsection{Primary Data Reduction}

The primary reduction of the dataset was carried out using the standard MEGARA 
data reduction pipeline \citep[DRP; ][]{2018zndo...2206856P}. As described 
above, MEGARA/IFU data consist of 623 spectra corresponding to 567 object and 56 
sky fibres, arranged into groups of multiple minibundles. In the spectral image, 
these groups are separated by gaps and have progressively smaller number of 
minibundles as they were built to reconstruct the curvature of the pseudo-slit 
at the focal plane of the MEGARA spectrograph. Flat-field images are used to
trace the locus of each of the 623 spectra using an automatic routine. 
The routine fits simultaneously 623 Gaussians every 200 columns and then 
interpolates the parameters of the Gaussian for each spectral pixel. 
With this information, the routine generates a weight map for every fibre that 
is applied to the data in order to perform the extraction. The procedure allows 
correcting the flux of each fibre for cross-talk contamination from 
adjacent fibres to a level of a few per cent, depending on the quality of the 
spectrograph focus during the observation and the shape of the spectral PSF 
for each wavelength and fibre.

A shift along the pseudo-slit axis of the trace locus with dome temperature 
has been noticed, which needs a correction of the locus for each observed frame. 
We used sky lines to measure interactively the shifts, which are found to be 
$\sim$2 pixels for the three spectral images used in this work. The final product of 
the DRP is a wavelength-calibrated, sky-subtracted 2D spectral image. 
This image contains 623 spectra, one spectrum corresponding to each fibre. 
The standard star is also reduced in an identical fashion.
Spectra of all fibres containing the standard star are summed to extract a 1D
spectrum, which is used to obtain the flux sensitivity curve using the 
{\sc iraf}{\footnote{{\sc iraf} is distributed by the National Optical Astronomy 
Observatories, which are operated by the Association of Universities for 
Research in Astronomy, Inc., under cooperative agreement with the National
    Science Foundation.}  } routines
for this purpose. The resulting sensitivity curve is used to obtain the
flux-calibrated 2D images in selected spectral lines and the 3D data cube. 

\begin{figure}
\begin{centering}
\includegraphics[trim=5mm 3.9cm 3mm 3.5cm, clip, width=1.05\linewidth]{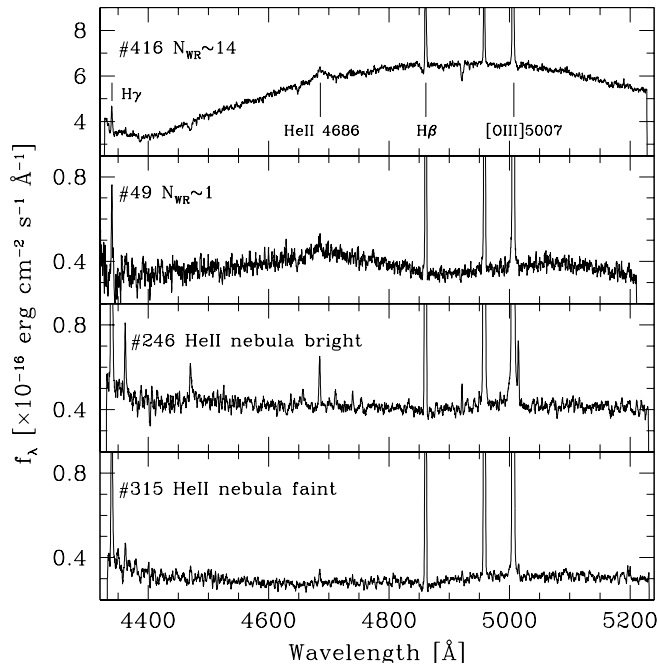}
\caption{
Spectra of individual fibres selected to illustrate broad emission at \heiiwr\ 
corresponding to WR stars (top two panels) and narrow \heiiwr\ from ionized nebula
(bottom two panels). Four emission features are marked in the top-most spectrum.
Fibre\#416, which belongs to SSC-A, illustrates the case of a spectrum containing 
multiple WR stars, whereas fibre\#49 illustrates the spectrum of a marginal 
detection of a WR star. Fibres\#246 and \#315 are examples of zones with bright 
and faint nebular \heiiwr\ line, respectively. 
}
\label{fig:spectra}
\end{centering}
\end{figure}

In Fig.~\ref{fig:spectra}, we show spectra of 4 fibres illustrating the 
detection of broad (top two panels) and narrow (bottom two panels) \heiiwr\ 
features. The Balmer lines H$\beta$ and H$\gamma$, and the 
[OIII]$\lambda$5007~\AA\ line are also indicated.

\subsection{Line maps, data cube and astrometry}

The positions of the fibres in the image plane with respect to the image 
centre are given in millimeters in the image headers. The 2D spectral image 
is converted into a 3D data cube using these fibre coordinates. A plate scale 
of 1.212~arcsec\,mm$^{-1}$ and a fibre diameter of 0.62~arcsec were used to 
transform the image coordinates from cartesian  to equatorial systems. 
To create a map at any sampled wavelength (or a selected band corresponding to a
line or a continuum region), we deposited the flux at that wavelength into 
hexagonal areas (the hexagonal spaxel is inscribed in a circle of 0.62~arcsec 
diameter), which correctly simulates the FoV of each fibre on the sky. 
Alternatively, to visualize extended emission, we also created smooth images 
by depositing the flux of a fibre into a highly oversampled pixel 
(size=0.021~arcsec) and then convolving it with Gaussian kernel of $\sigma$=half 
the fibre size (0.31~arcsec). Finally, the World Coordinate System header 
parameters were updated to force the RA and DEC of SSC-A to their values 
(RA=04:30:48.233, DEC=+64:50:58.59) in the GAIA-DR2 system. After creating the 
image, we measured a mean coordinate error of $\sim$0.3~arcsec for different 
sources in the resulting image.

We used the QFitsView facility \citep{Ott2012} for a quick visualization of 
the datacube and inspection of spectral features associated to known structures 
in the image. However, all analysis were carried out on the spectrum of an 
individual fibre or a sum of spectra of neighbouring fibres associated to a 
physical source. A datacube from the 2D spectral data was constructed using 
{\sc{megararss2cube}}\footnote{{\sc megararss2cube} is a tool written in python 
to convert MEGARA reduced dataproducts from the RSS format obtained with 
megaradrp to a more user-friendly 3-D datacube,  available as a repository in 
GitHub: \href{https://github.com/javierzaragoza/megararss2cube}
{https://github.com/javierzaragoza/megararss2cube}}, 
whereas the generation of line maps and their astrometry, extraction of 
spectra for physical regions and measurement of line fluxes were all carried 
out using our own scripts\footnote{MEGARA-related {\sc iraf} scripts are available 
on request to the first author.} in the {\sc iraf} environment.

We calculated the error as the 1-$\sigma$ deviation, $\sigma_l$, on each measured 
line flux using the expression \citep{Tresse1999}: \\
\begin{equation}
\sigma_l = \sigma_c D \sqrt{(2 N_{\rm pix} + \frac{EW}{D})},
\end{equation}
where $D$ is the spectral dispersion in \AA\ per pixel, $\sigma_c$ is the mean
standard deviation per pixel of the continuum, $N_{\rm pix}$ is the number of
pixels covered by the line, and EW is the equivalent width of the measured 
line. We used the value of FWHM to substitute for $N_{\rm pix}$.

\begin{figure*}
\begin{centering}
\includegraphics[width=0.325\linewidth]{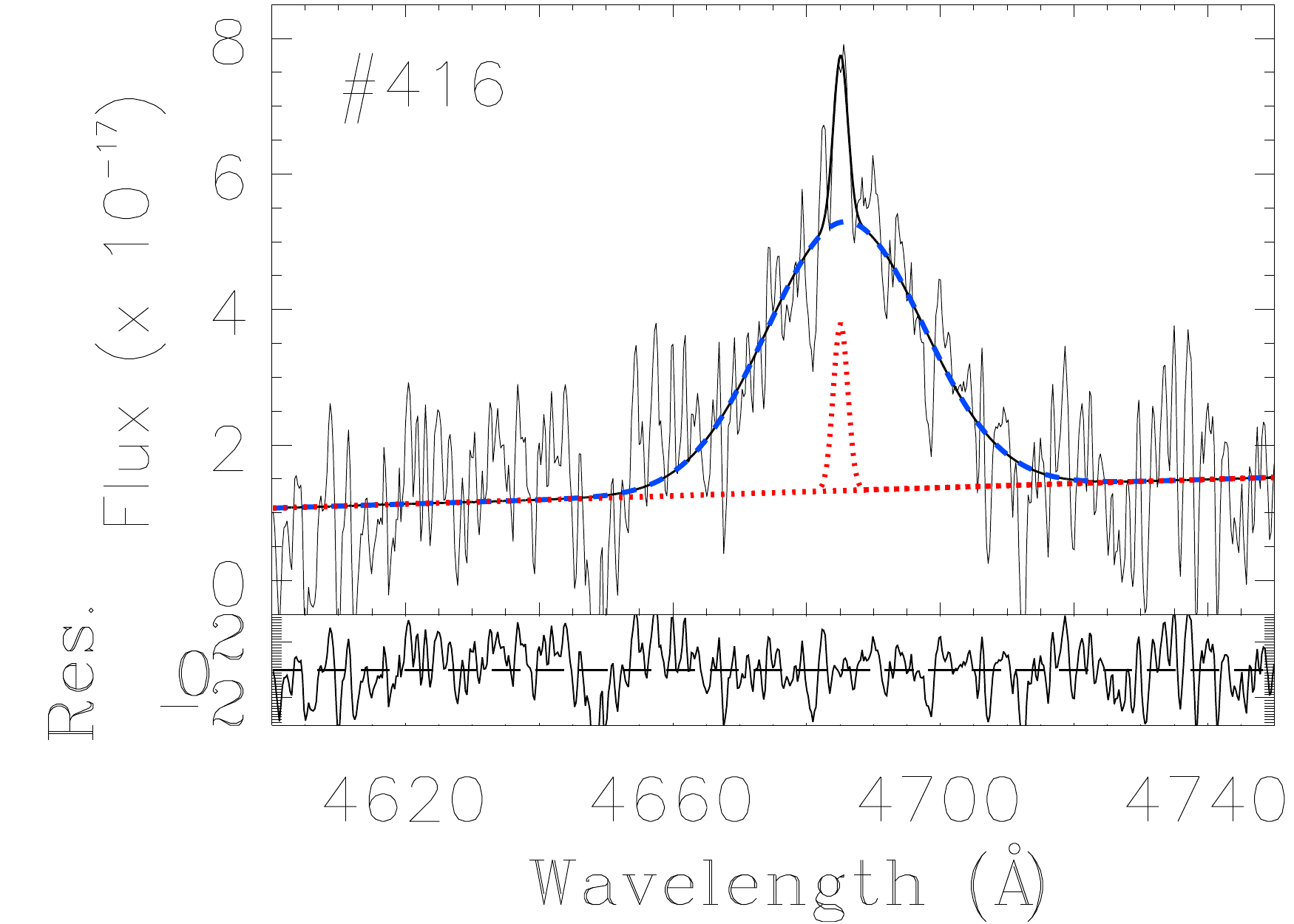}~
\includegraphics[width=0.325\linewidth]{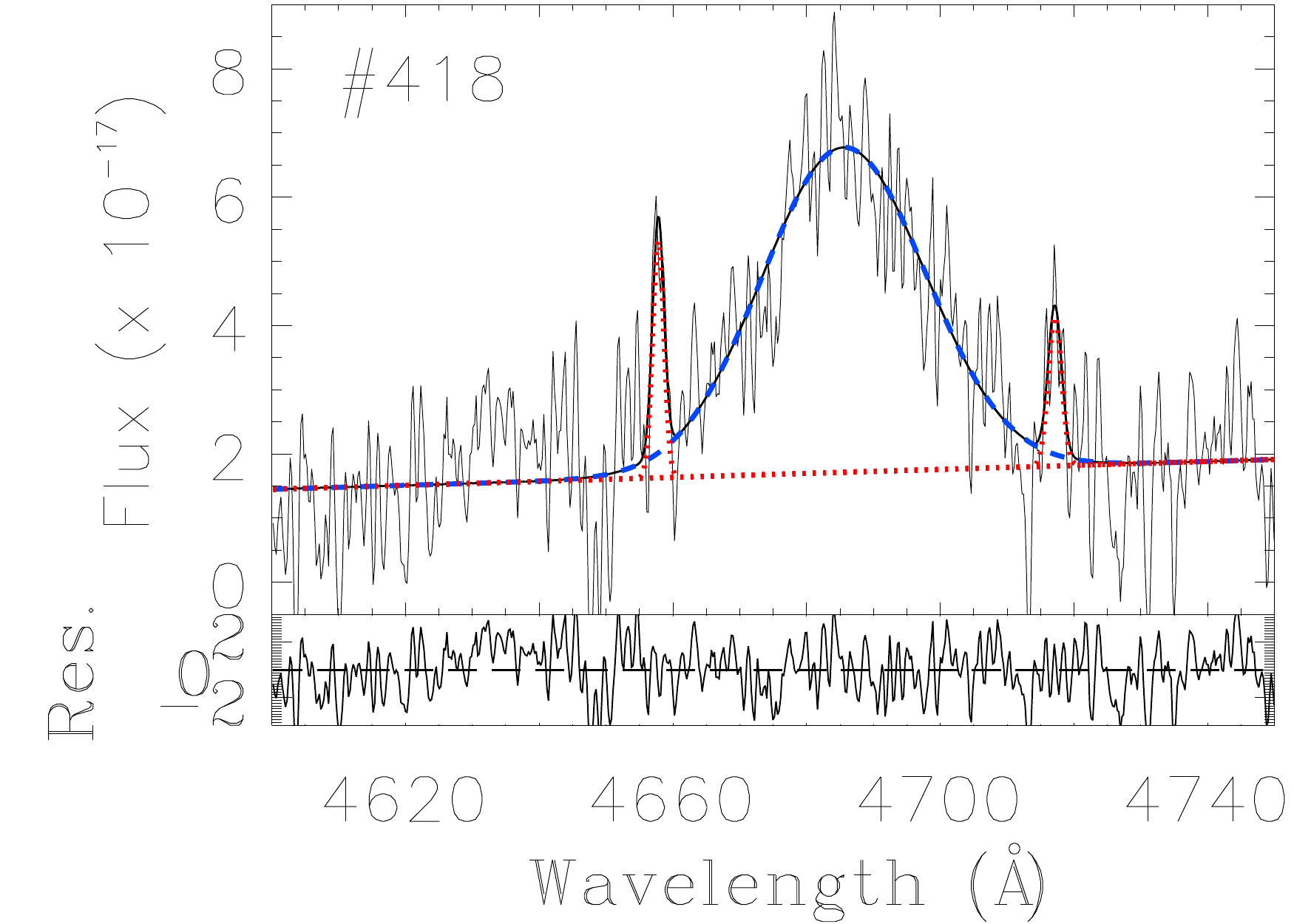}~
\includegraphics[width=0.325\linewidth]{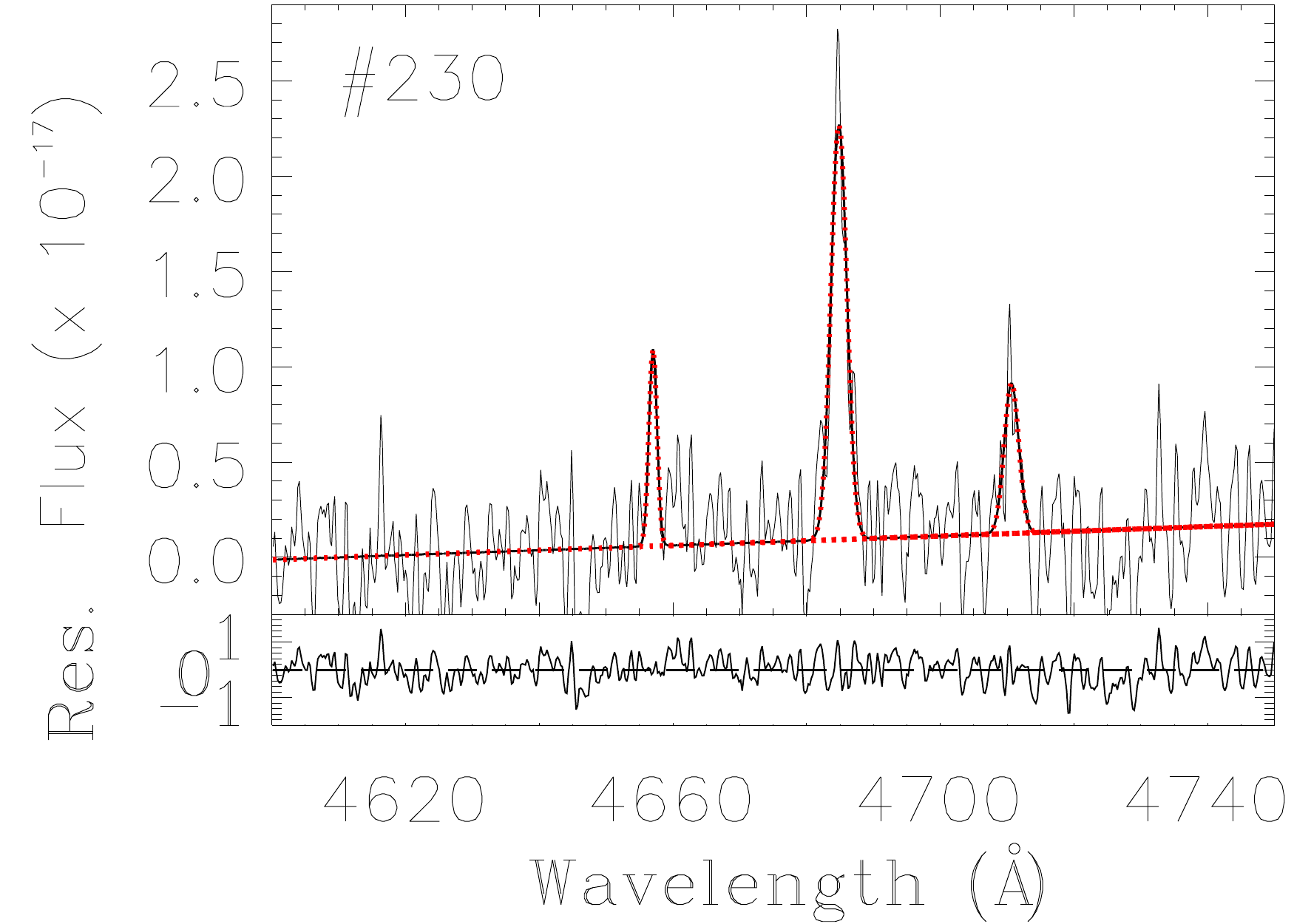}
\caption{Multi-Gaussian fits to the blue bumps of individual fibres: 
broad \heiiwr\ line (blue), narrow \heiiwr\ nebular line (red) 
are shown for 3 representative fibre spectra that require: both the broad 
and narrow components (left; fibre\#416), only the broad component 
(middle; fibre\#418), and only the narrow component (right; fibre\#230). 
The fitted continuum is shown by the red straight line, and the 
residuals of the fit are shown below each fit. The fluxes in the 
fitted spectra as well as in the residual are in units of $10^{-17}$~\ergcmsang.
The middle and right spectra show additional narrow lines that we attribute to
\feiii\ and \hei$\lambda$4713 nebular lines.
}
\label{fig:guassian_fits}
\end{centering}
\end{figure*}

\subsection{Ancillary Data}

We used \hst\ images in F555W (ACS/HRC), F814W (ACS/WFC), F658N (ACS/WFC) and 
F469N (WFPC2) bands to associate our spectra to known structures in the image. 
All the \hst\ images were brought to GAIA-DR2 coordinate system using the GAIA 
stars in the \hst\ images. The first two images allow us to locate the stars and 
clusters, whereas the latter two trace the \ha\ and \heiiwr\ emitting sources, 
respectively. We used the identifications of the SSCs by \citet{Hunter2000} and
WR and nebular \heii\ sources by \citet{Buckalew2000}. We also used the 
Chandra/ACIS X-ray image in the 0.2--10~keV band from \citet{Monica2015} in 
order to locate the point and diffuse X-ray sources with respect to the optical 
sources. 

\section{Detection of WR and nebular \heiiwr\ emission}

A simple visual inspection of the 2-D spectral image suggested the presence
of a feature around 4686~\AA\ in many fibre spectra. Gaussian profile fitting 
of this feature suggested that in most cases, the feature is narrow with 
FWHM$\sim$1--2~\AA, comparable or slightly above the resolution of the 
spectrograph. In some other fibres, the feature is broad with FWHM$>$6~\AA.
These narrow and broad components are illustrated in the bottom two and top 
two spectra in Fig.~\ref{fig:spectra}, respectively. The narrow feature 
is easily identified as the \heiiwr\ emission line from the ionized nebula,
whereas the broad component is identified as the blue bump (BB), which is 
the distinguishing characteristic of WR stars. The BB is a broad spectral 
feature between 4600 to 4700~\AA, and consists of broad lines of He and/or N 
ions from nitrogen-rich WR stars (WN-types), or He and/or C ions from carbon-rich 
WR stars (WC-types). Among these, the prominent N and C lines are \niiiwr, \nvwr, \ciiiwrb\ 
and \civwrb. 
WC stars, and hence lines from C ions, are generally absent in low-metallicity 
environments \citep[see e.g.~figure~8 in][]{Esteban2010}.
Additionally, nebular lines \heineb, \heiiwr\ and \feiii\ can also 
contribute to the BB. Some spectra showed both narrow and broad \heiiwr\ 
features, which required an analysis using multi-Gaussian fitting 
for the recovery of each component.

Another factor that affects the measurement of fluxes of relatively faint lines 
is determining the precise shape of the continuum, which requires a signal-to-noise 
ratio (SNR) of at least 10 in the continuum on either side within $\sim$100~\AA\ 
of the line of interest. In most continuum-weak 
fibres this condition is not met. For example, the SNR of the continuum in the 
bottom three spectra in Fig.~\ref{fig:spectra} is $\sim$5 at 4400~\AA\ to 
$\sim$9 at 4800~\AA. On the other hand, the top spectrum, which belongs to
SSC-A, has a minimum SNR of 23 at 4400~\AA, increasing to 43 at 
4800~\AA. Stellar absorption lines, most of which are reported by 
\citet{Rosa1997} for SSC-A, can be seen in the top-most spectrum.

\subsection{Multi-component Gaussian decomposition of the blue bump}

It is well established that the individual lines that contribute to the BB
can be extracted from multi-component Gaussian decomposition technique 
\citep[see e.g.][]{Brinchmann2008}. We hence carried out multi-Gaussian 
decomposition fittings using a custom-made code which uses the {\sc idl}
routine {\sc lmfit}\footnote{The {\sc lmfit} function (lmfit.pro) does
  a non-linear least squares fit to a function with an arbitrary
  number of parameters. It uses the Levenberg-Marquardt algorithm, 
incorporated in the routine {\it mrqmin} of Numerical Recipes in C 
\citep{Press1992}.}\citep[see][]{Gomez-Gonzalez2020}. 
Before fitting multiple Gaussians, a continuum level is defined for each 
spectrum in a two-step procedure. First, a large-scale continuum is defined 
for each spectrum using the task {\it continuum} in {\sc iraf}, by fitting 
a high-order polynomial (spline3, order=11) passing through carefully chosen 
line-free bins in the entire wavelength range of the observed spectrum. 
Second, any residual local continuum around the BB is accounted for by linearly 
interpolating line-free zones on either side of the BB. Two of the 3 parameters, 
the peak intensity $I_0$ and the line width $\sigma$ that define each Gaussian, 
were left free. Any line with a FWHM$>6$~\AA\ is defined as a broad line 
associated to a WR star. The third parameter, $\lambda_0$, is assigned to the 
rest wavelength of one of the expected lines from WR stars. The fitting program is 
executed interactively, where the bright nebular lines are fitted first, 
followed by the \heii\ broad component. Residuals are examined for a peak near 
any of the expected line wavelength. If present, a second broad line is fitted
and residuals are re-examined.  While fitting this second line, the $\sigma$ 
and $I_0$ of the first line were left free. The process continues examining the 
residuals and adding a new line until the residual flux is less than 3 times 
the root-mean-square (RMS) noise of the spectrum. In the iterative process, any 
faint nebular lines are fitted, if needed. The method is able to recover the 
\heiiwr\ components from WR (broad) and nebula (narrow) even when both are 
present in a single spectrum.
Fig.~\ref{fig:guassian_fits}  illustrates the multi-Gaussian fits for spectra 
from 3 individual fibres that require both a broad and a narrow \heiiwr\ line
(left), only a broad line (middle) and only the narrow line (right).
None of our fibre spectra showed broad lines from nitrogen or carbon ions.
A feature is recovered in many fibres at $\lambda$=4658~\AA\ (see the spectra in 
the middle and right panels), which matches the wavelength of the \civwrb\ 
(broad) feature. However, the width of this line is comparable to that of nebular 
lines, and hence we identify this feature with the nebular (narrow) line \feiii.

We used the fitting program to automatically identify broad and narrow 
\heiiwr\ features in each of the 567 spectra. As mentioned above, the flux at 
the peak of the fitted Gaussian should be at least 3 times the noise level for 
it to be considered a detection.
The narrow component, which is of nebular origin, 
is detected in 262 of these fibres and the broad component, in 50. Twenty 
five of the spectra showing a broad component have an associated narrow 
component, whereas the remaining 25 show only a broad component. 

\begin{figure*}
\begin{centering}
\includegraphics[width=0.8\linewidth]{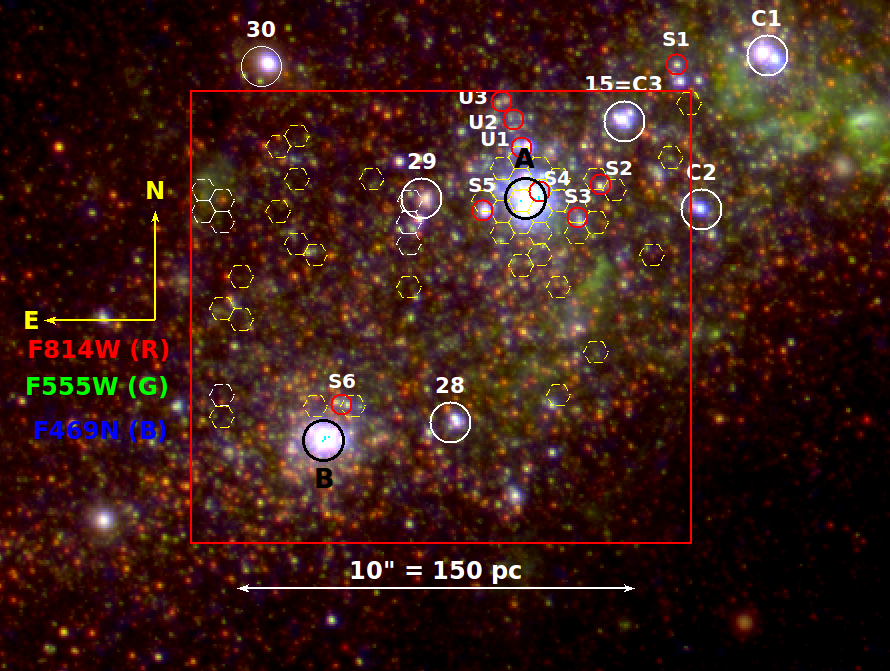}
\caption{
A colour-composite RGB image of the central starburst zone of NGC\,1569.
The image is formed by combining \hst\ F814W, F555W and F469N band images, as red, 
green and blue components, respectively. MEGARA/IFU FoV 
(12.5~arcsec$\times$11.3~arcsec; red rectangle) and image orientation (compass) 
are shown. The locations of the fibres where we detected a broad \heiiwr\ 
component are shown by hexagons. The prominent objects in the FoV are identified. 
These include the two massive clusters SSC-A and SSC-B, and four other clusters 
(15, 28, 29 and 30) from \citet{Hunter2000}. WR sources identified by 
\citet{Buckalew2000} using the displayed \hst\ F469N image, which intercepts the 
\heiiwr\ line, are shown. These include four clusters with WR stars (C1, C2, C3 
and C4) and six WR stars (S1, S2, S3, S4, S5 and S6). Three sources (U1, U2 and 
U3) classified as of unknown origin by \citet{Buckalew2000} are also shown.
The \heiiwr\ broad feature is detected in 18 adjacent fibres associated to SSC-A.
The circles enclosing cluster candidates 
have a diameter of 1~arcsec, and hexagons have an equivalent diameter of 0.62~arcsec.
}
\label{fig:wrimages}
\end{centering}
\end{figure*}

\begin{figure*}
\begin{center}
\includegraphics[scale=0.325]{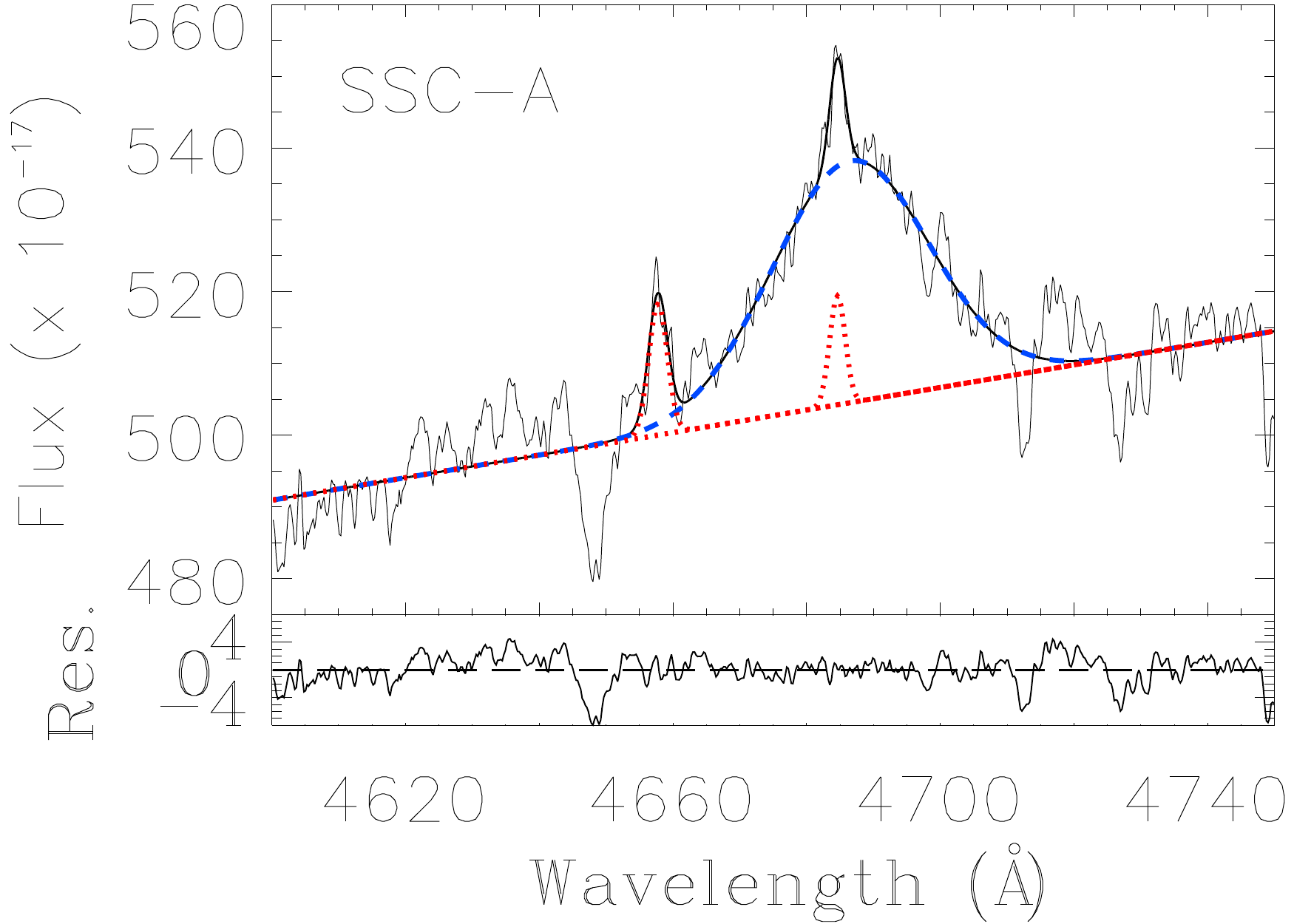}~
\includegraphics[scale=0.325]{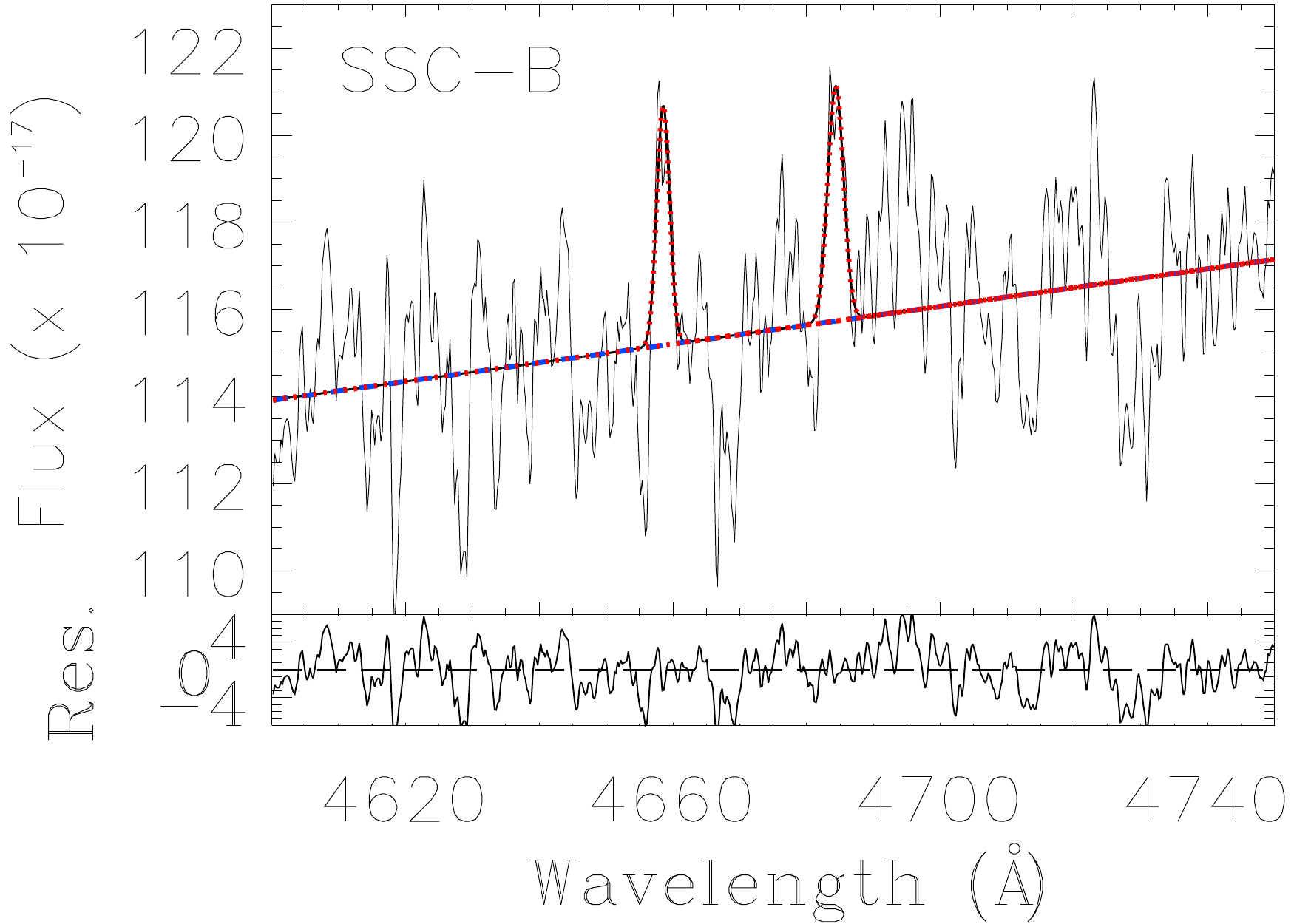}
\includegraphics[scale=0.325]{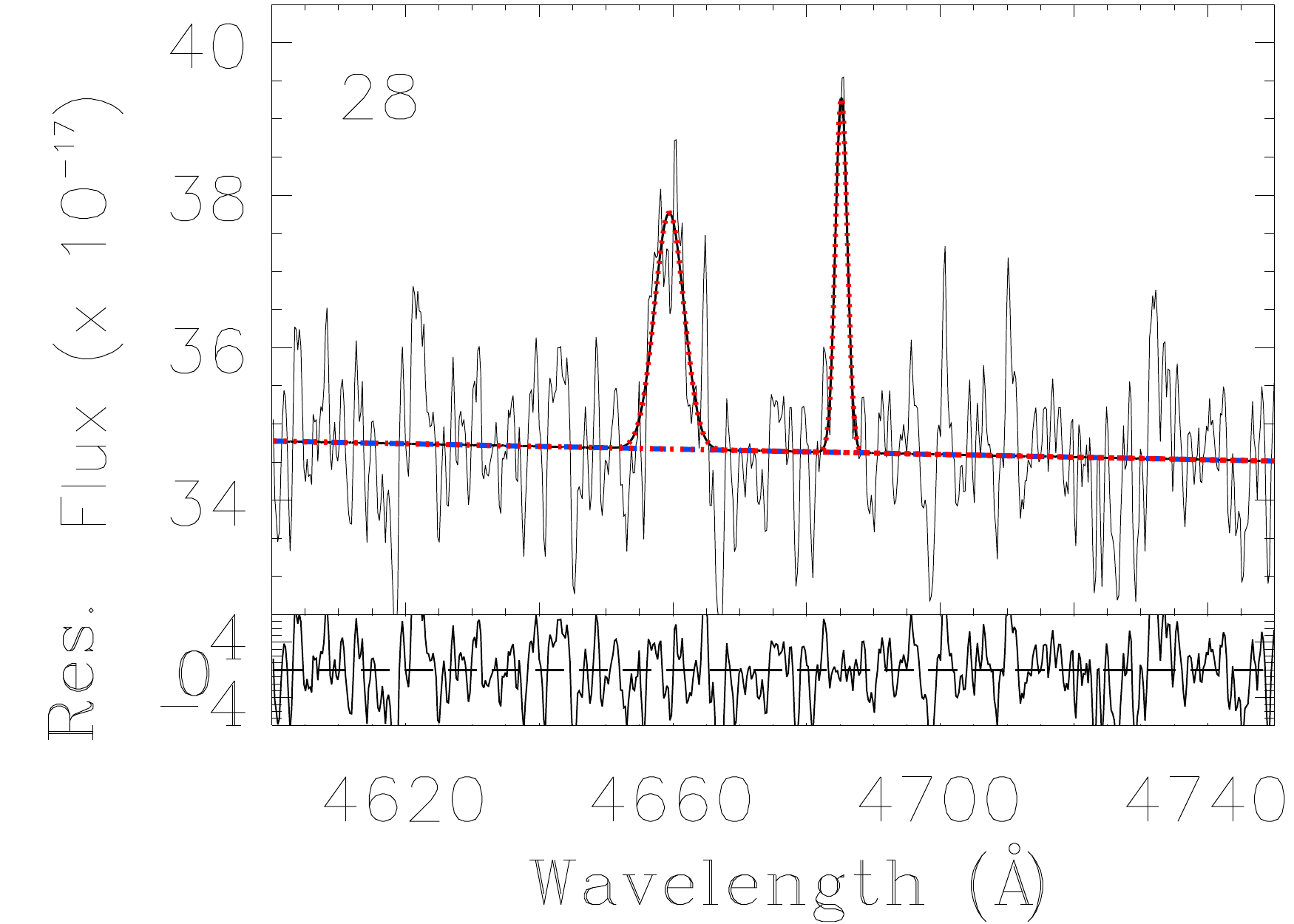}\\
\includegraphics[scale=0.325]{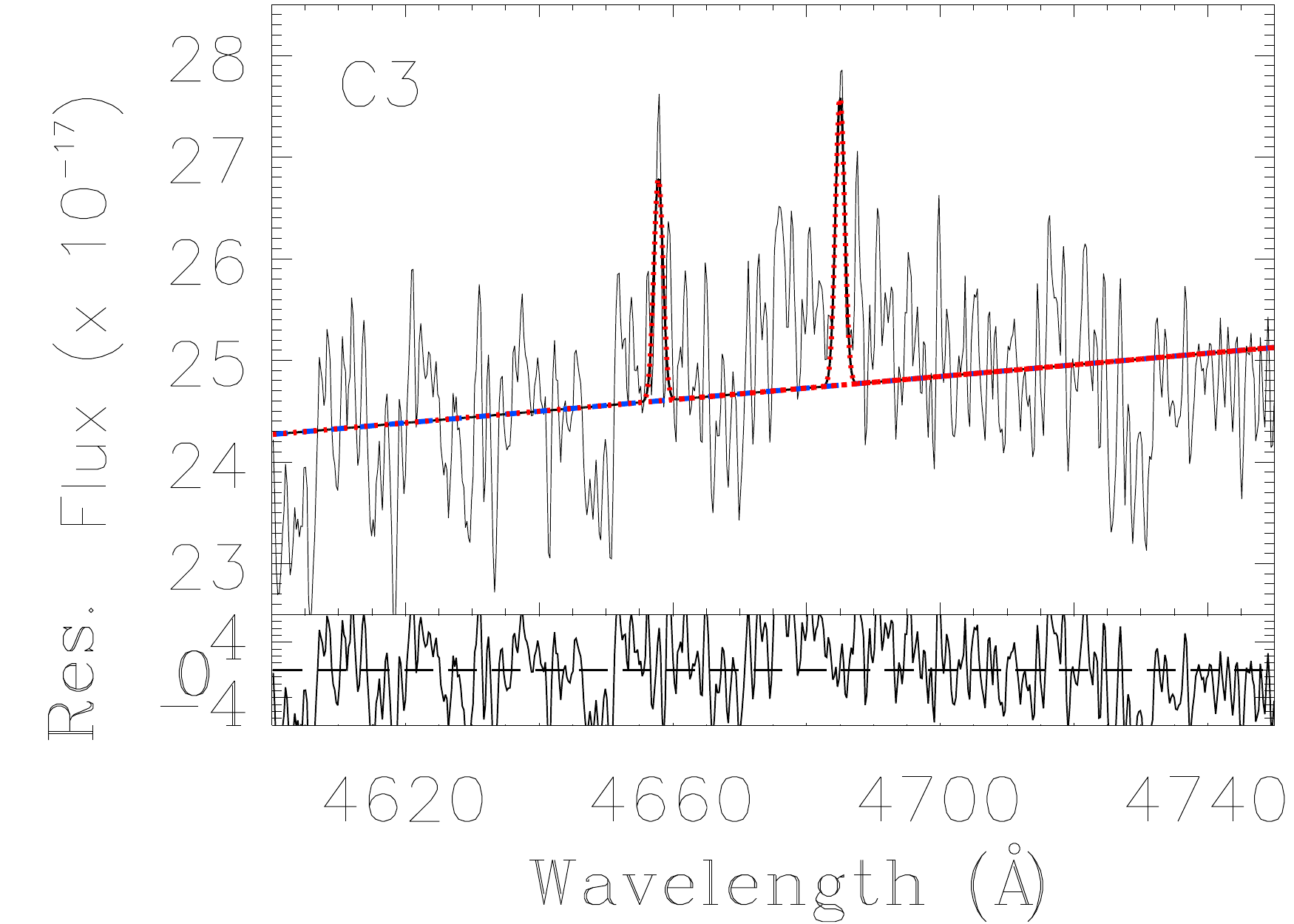}~
\includegraphics[scale=0.325]{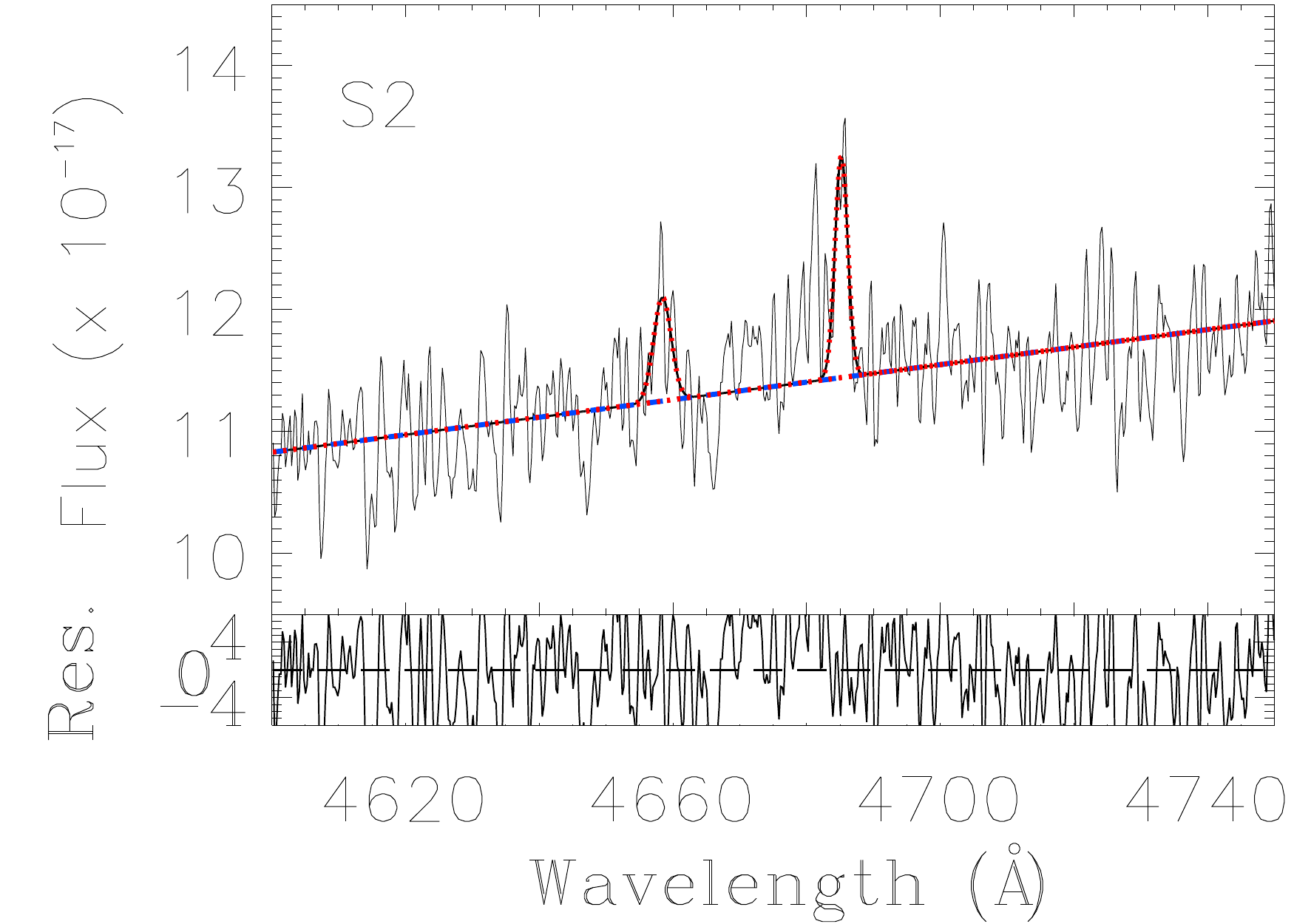}
\includegraphics[scale=0.325]{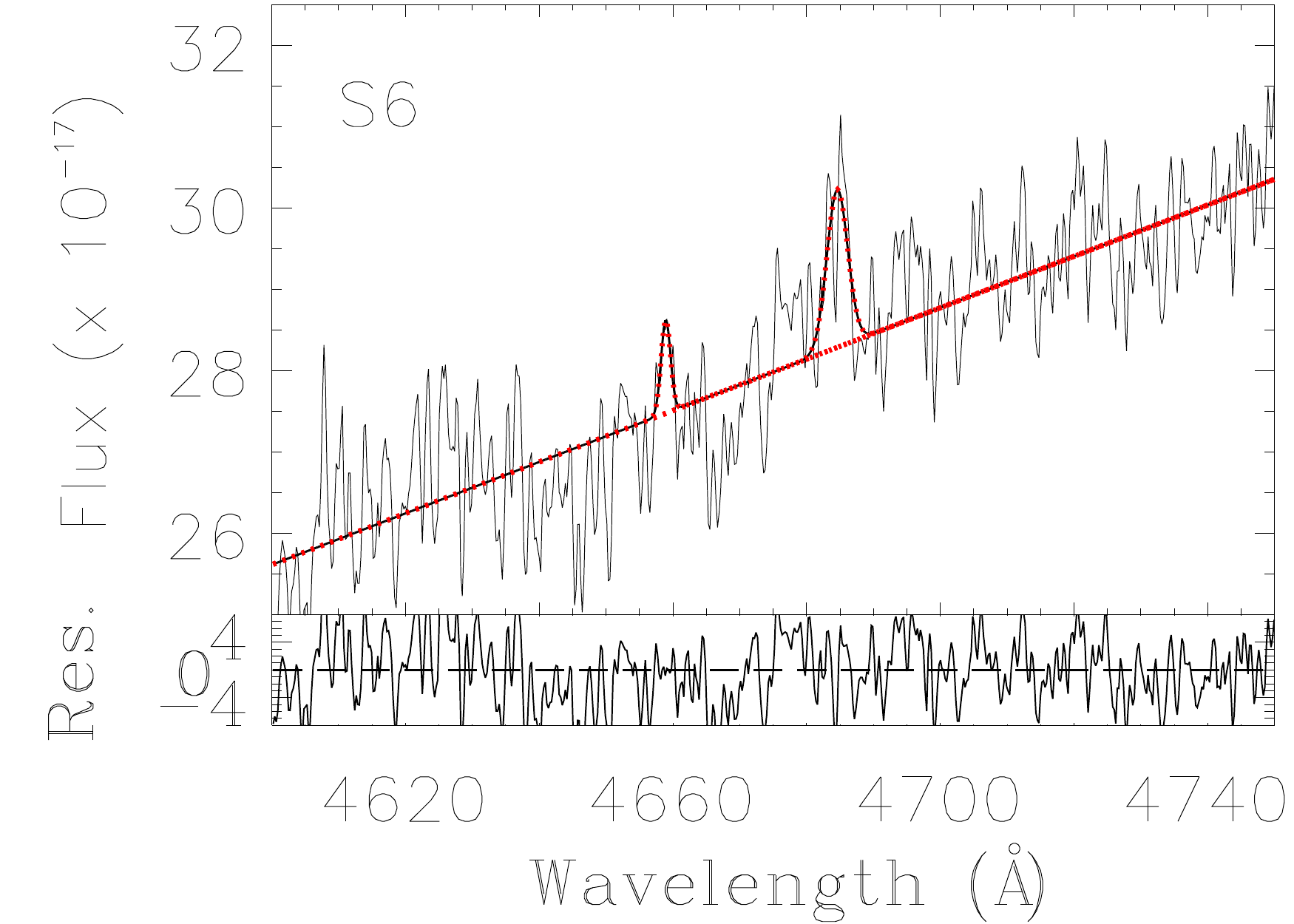}
\caption{
Multi-fibre summed spectra around the BB of SSC-A and other cluster and stellar 
sources in the FoV of MEGARA/IFU. No WR stars are detected in clusters SSC-B 
and 28. Star cluster C3 and stellar sources S2 and S6, where the presence of
WR stars have been previously reported from the analysis of the \hst\ F469N-band 
narrow-band images, only show nebular lines in our spectra. The fluxes are 
plotted in units of $10^{-17}$~\ergcmsang, and the bottom panel of each plot 
shows the residual flux in the same units.
}
\label{fig:wr_candidates}
\end{center}
\end{figure*}

\subsection{Location of fibres with WR features}

In Fig.~\ref{fig:wrimages}, we show the positions of all the 50 individual 
fibres with broad \heiiwr\ detections 
(hexagons\footnote{See Fig.~\ref{fig:map_qla} in the appendix for the 
identification numbers of each fibre.}), 
overlaid on an RGB colour-composite image formed by \hst\ images in F814W, F555W 
and F469N filters, shown as red, green and blue colours, respectively.  
This set of images was chosen in order to highlight the candidate
WR stars, which are expected to be continuum-bright sources with 
\heiiwr-excess. The red and green images having been 
taken using broad bands, trace continuum sources, whereas the blue image traces 
sources that have contribution from emission lines in the 4650--4720~\AA\ range.
The \heiiwr\ line is the most likely contributor to this image, but as the spectra in 
Fig.~\ref{fig:guassian_fits} illustrate, nebular lines \feiii\ and 
\hei\,$\lambda$4713 could also contribute in this filter. Thus, blueish-looking
continuum-bright sources in this image are the likely WR candidates.
\citet{Buckalew2000} used the F469N image in combination with continuum
and nebular images to identify WR candidate sources. They classified these as 
cluster, stellar, and unknown origin sources, and named them as
C1 to C4, S1 to S6, and U1 to U3 sources, respectively,
which are all identified in the figure. In addition, we identified stellar 
clusters catalogued by \citet{Hunter2000} (numbers 15, 28, 29 and 30) and the 
two well-known SSCs A and B in the figure. Cluster C4, which lies just outside
our FoV and is identified as cluster 10 by \citet{Hunter2000}, is the ionizing
cluster of the brightest \hii\ region in this galaxy.

The presence of the \heiiwr\  broad component is the characteristic
signature of a WR star. However, given that our observations were carried
out at $\sim$0.9~arcsec seeing and the fibres cover a diameter of 0.62~arcsec,
several adjacent fibre spectra should also show up the broad feature in order
to associate the inferred feature with the detection of a WR star.
Out of the 50 fibres where we inferred \heiiwr\  broad component,
18 contiguous fibres are associated to SSC-A. Some of these fibres could be 
associated to candidate stellar sources S3, S4 and S5, which are located within 
the seeing-convolved image of SSC-A. 

Of the remaining detections, we infer two locations with four (\#417, 419, 
421, 423; on the top-left) and three (\#145, 149, 153; on the top-middle) 
contiguous fibres, associated to them. 
However, a careful scrutiny of the fibres associated 
to these locations revealed that these fibres are physically next to one of the 
18 fibres belonging to SSC-A on the psuedo-slit, and hence the detected broad 
bump is likely arising due to residual cross-talk from SSC-A. Hence, we ignore 
the detections in these seven fibres. 
The rest of the inferred \heiiwr\  broad component corresponds to 25 single-fibre 
detections, scattered all over the image. One of these detections (\#413) is due 
to cross-talk with SSC-A.
Of the remaining detections, two coincide with the previously reported candidate 
stellar sources S2 and S6. If these single-fibre detections come from real 
sources, spectra extracted by summing spectra of fibres adjacent to the location 
of these fibres also should show a broad \heiiwr\ feature. We carried out the 
Gaussian decomposition of the BB with the specific purpose of detecting these 
broad features in spectra obtained by summing spectra of at least 3 fibres 
around the one where we detected the broad feature. None of these summed spectra 
around the 25 single-fibre detections showed a broad component. Uncertainty in 
defining the continuum is the likely reason for the feature to vanish in summed 
spectra in spite of them being detected, i.e. the peak flux of the fitted broad 
(FWHM$>$6~\AA) Gaussian profile is $>$3 times the RMS error of the continuum, in 
individual fibre spectra. The spectrum for fibre \#49 in Fig.~\ref{fig:spectra} 
is one such example. Deeper observations would be required to ascertain the 
nature of these sources. We consider these detections as marginal and in \S4.3 
we compare their fluxes to that expected from typical WNL-type WR stars.

\begin{figure*}
\begin{centering}
\includegraphics[width=0.8\linewidth]{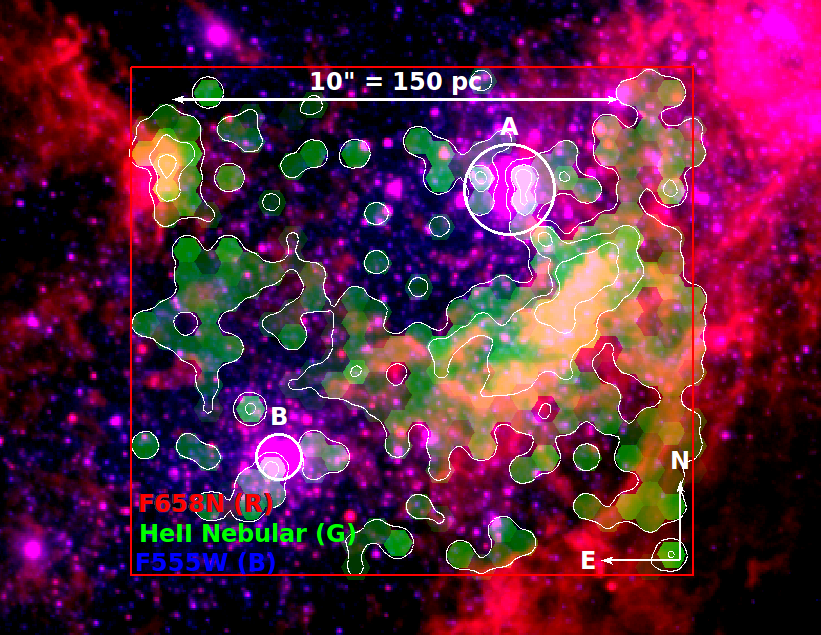}
\caption{
Colour-composite RGB image showing the morphology of the \heiiwr\ nebular 
emission (white contours englobing green hexagons) with respect to that of the 
\ha\ emission (\hst\ F658N-band shown in red) and stellar continuum (\hst\ F555W 
image shown in blue). Clusters A and B are marked. The contours corresponds to 
surface brightness levels of 3, 12 and 20$\times10^{-17}$~\ergcmsarc\ 
in the \heiiwr\ nebular line. The nebular emission is weak from 
SSCs A and B, and is distributed along a semi-circular arc with the most 
intense emission coming from a zone $\sim$40~pc south-west of SSC-A.
}
\label{fig:map_heiineb}
\end{centering}
\end{figure*}

The summed spectra allow us to understand the reasons why \citet{Buckalew2000} 
interpreted the observed excess emission in the \hst/F469N band in C3, S2 and S6
as originating in WR stars. It may be recalled that imaging observations cannot 
distinguish between the narrow and broad components. \citet{Buckalew2000} 
assumed it to be a WR detection, if the position of a star or cluster coincides 
with the position where an excess flux is detected in the F469N band. This 
selection criterion does not contemplate the presence of extended \heiiwr\ 
nebular emission. As we discuss in detail below, we have detected extended 
\heiiwr\ nebular emission in the central zone of NGC\,1569. Objects C3, S2 and 
S6 are stellar sources in which we detect narrow nebular \heiiwr\ and \feiii\ 
lines in our spectra (see the bottom three spectra in Fig.~\ref{fig:wr_candidates}). 
The narrow lines are not originating in the stellar 
sources, and are instead part of the extended nebular emission.
At the location of the three sources of unknown origin (U1, U2 and U3), we 
detect only narrow \heiiwr\ line; these sources are also part of the
extended nebula.

Our spectral data are used to infer the WR population in known clusters in our 
FoV. Apart from SSC-A, where we already demonstrated the presence of WR stars, 
our FoV includes clusters SSC-B, 28, 29, and C3. Among these, previous 
spectroscopic observations have already ruled out the presence of WR features 
in SSC-B \citep[e.g.][]{Rosa1997}. No WR stars were inferred in F469N images by 
\citet{Buckalew2000} from sources 28 and 29, whereas C3 is a candidate WR cluster. 
We extracted spectra around all these sources, as well as for SSC-A (C4), 
by adding spectra of 7 fibres around the location of these clusters. 
Unfortunately, cluster 29 spectrum suffers from cross-talk problem as several 
of its fibres are adjacent to some of the fibres of SSC-A on the pseudo-slit. 
We hence did not carry out an analysis of search for WR stars in this cluster.
Multi-Gaussian analysis was carried out on these summed spectra, which
are shown in Fig.~\ref{fig:wr_candidates}.
These fits establish the presence of WR features in summed spectra of
SSC-A, and their absence in SSC-B, 28 and C3. The figure also shows (last two 
panels) the results of the fits to the summed spectra for S2 and S6. Given 
the closeness of these sources to SSC-A and SSC-B, respectively, we summed 
spectra of only 3 adjacent fibres in these cases.  
The broad \heiiwr\ component is not recovered in these summed spectra.

\subsection{\heiiwr\ nebular morphology}

We now discuss the results obtained by muti-Gaussian fitting to recover the
\heiiwr\ nebular line. Unlike the broad component, the narrow \heiiwr\ line, 
which is of nebular origin, is detected in nearly half of the total fibres. 
The fibre fluxes and locations are used to create a map, which is shown in 
Fig.~\ref{fig:map_heiineb} as the green component of the RGB image. 
The \hst\ images in F658N (\ha\ + continuum) and F555W bands are used for the 
red and blue components, respectively. These latter images help us to see the 
large-scale distribution of ionized gas and stars at the \hst\ resolution. 
Contours corresponding to \heiiwr\ nebular line surface brightness of 
3, 12 and 20$\times10^{-17}$~\ergcmsarc\ are shown.

Clearly the \heiiwr\ narrow emission is not confined to a few point sources, 
but instead is part of an extended nebula. 
This emission is distributed along a semi-circular arc of 150~pc
(10~arcsec) diameter. The brightest part of the \heiiwr\ 
nebula lies to the south-west of SSC-A at a distance of $\sim$40~pc (2.5~arcsec).
This brightest part is also the zone closest to SSC-A, where there
is an increase in the \ha\ surface brightness, and corresponds to the \hii\ 
region numbered 3 by \cite{Waller1991}. The nebular arc is widest near 
this zone, reaching $\sim$40~pc width. The centre of the semi-circular arc
does not coincide with SSC-A, and instead is shifted to the east of it by $\sim$40~pc, 
roughly coincident with cluster 29. Unfortunately our FoV does not cover the 
northern part, hence we cannot conclude whether the observed arc is part 
of a complete circular nebula or not. There is no evidence for arc-like 
structure in the \ha\ image on the northern side, nor any of the ionized 
superbubbles (sb) identified by \citet{Monica2015} match our \heiiwr\ emitting 
segment or its possible northern counterpart. This can be due to the lower 
sensitivity of Fabry-P\`erot images presented by \citet{Monica2015} in 
comparison to the \hst\ images. The \heiiwr\ emitting segment lies between the 
superbubbles classified as sb4, sb5, sb6 and sb7. The \heiiwr\ nebular emission 
is weak in the immediate vicinity of SSC-A and SSC-B. 

By mapping  the \heiiwr\ nebular emission we can  obtain the total flux in this 
line, and from this, the ionization requirements. The data need to be corrected 
for extinction from dust along the line of sight in the Galaxy as well as
the dust in NGC\,1569, before we can carry out these calculations.
The availability of the extinction-sensitive Balmer lines \hb\ and \hg\ in our
dataset allows us to obtain the extinction from each fibre spectrum.
We describe the procedure adopted for extinction correction in the next section.

\section{Extinction-corrected number of WR stars and the \heii\ ionizing photon rate}

In this section, we present the extinction map of the zone covered by our IFU 
observations, and obtain the \heiiwr\ luminosity of the broad and narrow lines, 
corrected for extinction. The luminosity in the broad component is used to 
calculate the number of WR stars inferred from our observations, whereas the 
luminosity in the narrow component is used to obtain the He$^+$ ionizing photon 
rate. We used the emission line intensities of Balmer lines (\hb\ and \hg) 
in our IFU spectra to create maps of the ionized nebula and extinction. 
We note that the stellar continuum of unresolved populations, in general, 
experiences lesser amount of attenuation as compared to that experienced by 
the ionized gas \citep{Calzetti1994, Mayya1996}.
Hence, the number of WR stars derived after correcting the broad component 
luminosity for nebular extinction could be an over-estimate.

\subsection{Extinction map of the observed zone}

In \S3.1, we described the procedure we followed for decomposing the narrow 
and broad components of the \heiiwr\ feature. We measured the fluxes and 
related quantities of \hb\ and \hg\ nebular lines in each fibre spectrum using 
the single Gaussian fitting option of the IRAF task {\it splot} in batch 
mode. The line fluxes in each fibre are then transformed into maps using the 
procedure described in \S2.3.

\begin{figure}
\begin{centering}
\includegraphics[trim=1.5cm 6.5cm 1.0cm 7.5cm, clip, width=0.99\linewidth]{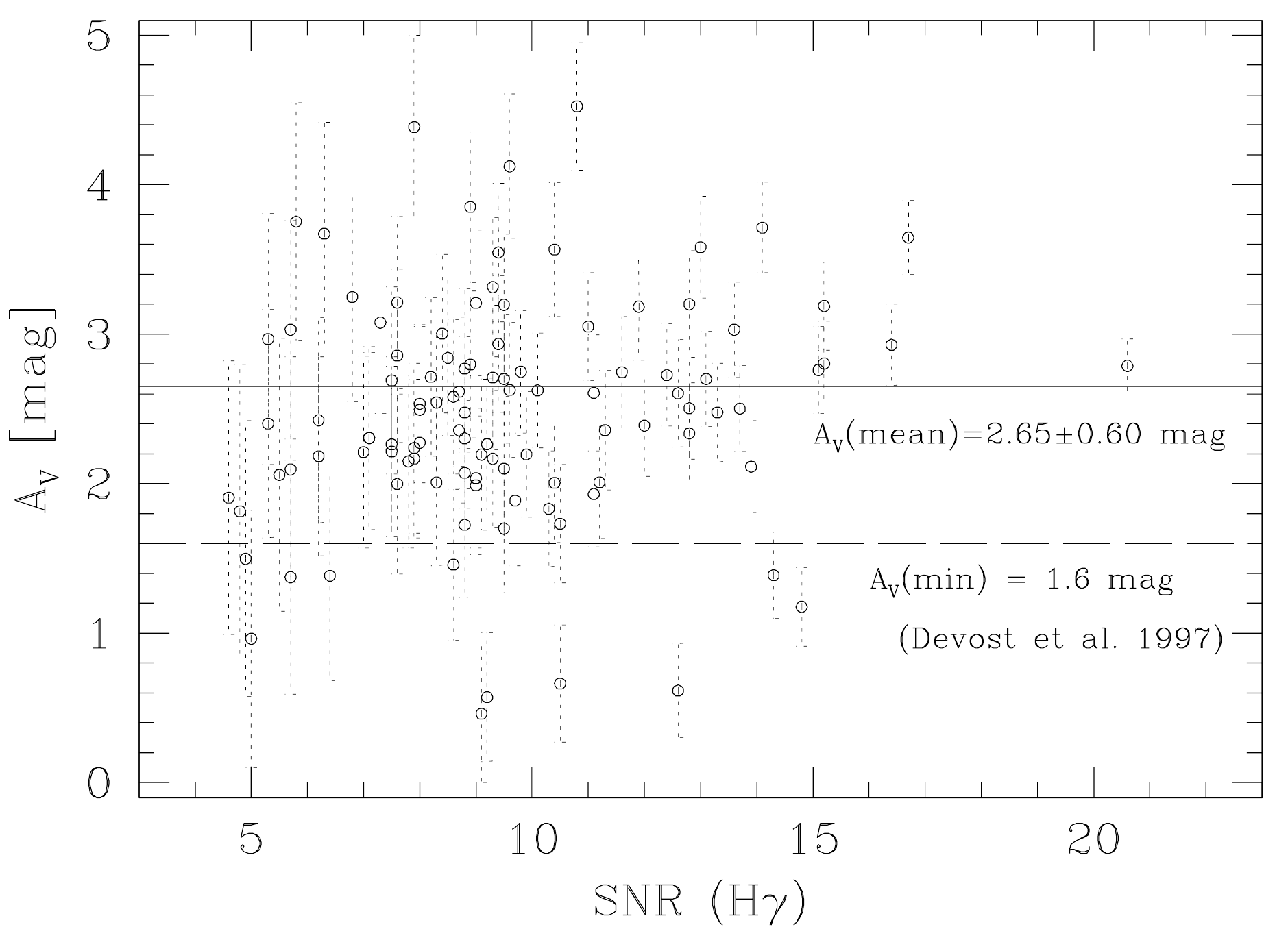}
\caption{
Visual extinction \Av\ obtained from the \hb\ and \hg\ nebular lines is plotted 
against the SNR of the H$\gamma$ line, which is the weaker of the two lines.
Only 106 spectra for which the EW(\hg)$>$8~\AA\ are considered for plotting.
Among the plotted points, 90\% lie above the dashed horizontal line, which 
corresponds to the minimum \Av\ in the entire galaxy found by \citet{Devost1997}. 
The mean value of our measurements is shown by the solid horizontal line. 
See \S4.1 for further details.
}
\label{fig:fig_av_vs_snr}
\end{centering}
\end{figure}

\hb\ and \hg\ emission fluxes were used to obtain the extinction from each 
fibre spectrum, where both of these lines were detected, using the Balmer 
decrement method for case B recombination of a typical photoionized nebula 
\citep[$T_{\rm e}$=10000~K, $n_{\rm e}$=100 cm$^{-3}$;][]{Osterbrock2006} 
and the reddening curve of \citet{Cardelli1989}. Only spectra with a SNR 
of at least 3 in both the lines are used. The \hb\ line is detected at 
SNR$>$7 in all the 567 fibres. This line is well-resolved with the FWHM of the 
fitted Gaussian varying between 1 and 2~\AA. Some spectra clearly showed signs 
of more than one velocity component. In 86\% of these spectra (487), the \hg\ 
line is also detected at SNR$\ge3$, enabling \Av\ measurements in all these 
spectra. The resulting values along with the errors are given in the last two 
columns of Table~\ref{table}. However, all measurements are not reliable.
A source of error in the derived \Av\ is the uncertain correction of the 
underlying absorption in \hb\ and \hg\ lines. We used a uniform 1~\AA\ of 
correction in EW proposed by \citet{Rosa1997} for both the \hb\ and \hg\ lines. 
The underlying absorption affects spectra at locations where we detected strong 
continuum, some of which belong to the WR stars. Thus, it is reasonable to use 
the \Av\ obtained for continuum-weak, or alternatively high EW, spectra for 
carrying out fibre-to-fibre \Av\ corrections. As \hg\ is the more critical of 
the two lines in \Av\ measurements, we used only those spectra that have 
EW(\hg)$>$8~\AA. This left us with the best 106 measurements of \Av.
The resulting \Av\ values are plotted against the SNR of the \hg\ line, 
which is the fainter of the two lines used for determining extinction, in 
Fig.~\ref{fig:fig_av_vs_snr}. The estimated error on each measurement is shown, 
which is typically $\sim$0.5~mag. These error bars do not take into
account errors introduced by the uncertain correction for the underlying
stellar absorption, e.g. the \Av\ values would be $\sim$0.2~mag higher 
or lower depending on no correction, or correction of 2~\AA\ (instead of the 
assumed 1~\AA), respectively.

\citet{Devost1997} obtained \Av\ using \ha\ and \hb\ lines for 16 
zones spread over the entire galaxy. They found all zones have \Av$\ge$1.6~mag,
with the minimum value corresponding to a zone in an ionized bubble, which is 
outside our FoV. They proposed that this minimum value corresponds to the 
extinction from the Galactic dust along the line of sight to NGC\,1569. 
\citet{Grocholski2012} obtained a marginally higher value (\Av=1.8~mag) using an analysis 
of the red giant branch of the resolved stellar population in the outer disk of the galaxy.
There are many independent measurements of the Galactic extinction in the 
direction of NGC\,1569. Using the line of sight HI observations, 
\citet{Burstein1984} and \citet{Schlegel1998} obtained values of 
$A_{\rm B}$=2.03~mag (equivalent to \Av=1.55~mag) and \Av=2.3~mag, respectively.
\citet{Israel1988} and \citet{Origlia2001} analysed the UV spectra of NGC\,1569
and suggested a foreground extinction of \Av=1.7~mag. More recently, 
\citet{Schlafly2011} obtained a value of \Av=1.9~mag based on the colour excess 
of Sloan Digital Sky Survey (SDSS) stars that have spectroscopic data.

\begin{figure}
\begin{centering}
\includegraphics[width=0.95\linewidth]{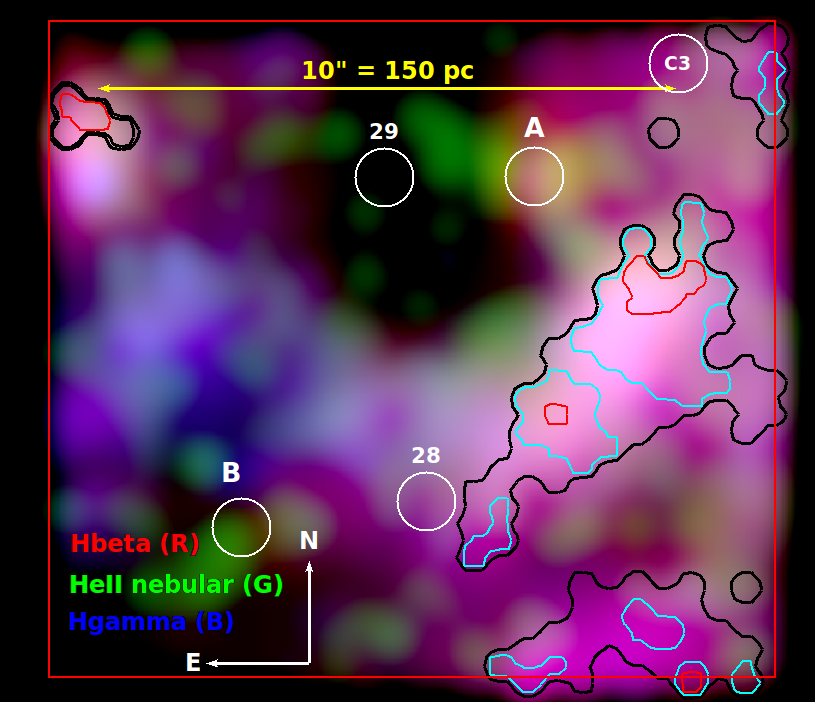}
\caption{
Extinction contours superposed on the MEGARA RGB nebular images.
Smoothed images in \hb, \heiiwr\ and \hg\ are used as red, green and blue 
components, respectively. The contours corresponding to \Av=1.6, 2.6 and 3.6~mag are shown 
in black, cyan and red colours, respectively. The noise in the measured fluxes
in individual fiber spectra did not allow the determination of reliable \Av\ 
values outside the black contours, though emission is seen in this smoothed image. 
Known sources in the FoV are labelled. The zone with the highest 
\heiiwr\ nebular emission (south-west of SSC-A) is also the zone with the highest \Av. 
The nebular emission follows a semi-circular arc of $\sim$150~pc diameter.
}
\label{fig:map_Av}
\end{centering}
\end{figure}

All these values, with the exception of one \citep{Schlegel1998}, are compatible 
with a Galactic extinction value of \Av=1.6~mag proposed by \citet{Devost1997} 
after taking into account measurement errors. We show this lower limit by a 
dashed line in Fig.~\ref{fig:fig_av_vs_snr}. Ninety per cent of our 
measurements are above this line. Nine of the 11 measurements below this line 
are consistent with \Av=1.6~mag if we calculate \Av\ without any absorption 
correction. Thus, taking into account external errors, our data are in agreement 
with a minimum \Av=1.6~mag proposed by \citet{Devost1997}. The rest of the \hii\ 
regions observed by \citet{Devost1997} have a mean value of $A_{\rm V}$=2.26~mag, 
implying an average internal extinction of 0.66~mag for NGC\,1569. We obtain a 
mean \Av=2.65$\pm$0.60~mag for 95 spectra having \Av$>$1.6~mag. This implies an 
average internal extinction of 1.05~mag for the zone covered by our observations.

We used the \Av\ values obtained in these 95 fibres to create an extinction map, 
whose contours are shown in Fig.~\ref{fig:map_Av}, overlaid on 
a nebular colour-composite image, formed by combining \hb, \heiiwr\ and \hg\ 
images as red, green and blue components, respectively. All the three nebular images
are smoothed using a Gaussian kernel of 0.31~arcsec following the procedure
described in \S2.3. As the \hg\ to \hb\ ratio is sensitive to extinction, 
the colour in the image is expected to change from blue to red as extinction increases.
The zone with the highest \heiiwr\ nebular surface 
brightness (south-west of SSC-A) is also the zone with the highest \Av, reaching 
values $\sim$4~mag. Extinction smoothly decreases from this point towards the 
south-east, reaching the Galactic values of \Av=1.6~mag at the boundaries of the 
detected zone. High extinction is also inferred from the zone with weak nebular 
emission in the bottom-right corner of the image, and a zone at the top-left 
corner of the image. Nebular lines are weak in almost all continuum-bright 
regions, including in SSC-A, which prevented us from determining \Av\ 
for these sources. We have assigned the Galactic values of \Av=1.6~mag to all 
regions outside the black contours, where \Av\ could not be reliably measured.
Fig.~\ref{fig:map_Av} also allows us to compare the large-scale morphology of \heiiwr\ 
nebula (green) with that of the ionized gas traced by Balmer lines at the same resolution.
The \hii\ gas is also distributed along the semi-circular arc traced by the 
\heii\ nebula, both showing a clear hole in the central zone of
their intensity distributions. 

\subsection{Total \heii\ ionizing photon rate}

The availability of an extinction map allows us to apply fibre-to-fibre 
extinction corrections. We used a uniform value of \Av=1.6~mag to all spectra 
where \Av\ could not be measured reliably. Summing all these fluxes, we obtain 
a total \heiiwr\ flux of $9.13\times10^{-14}$~\ergcms. This corresponds to a 
L(\heiiwr) of 
$1.0\times10^{38}$~\ergs\ and a Q(He$^+$) of $1.0\times10^{50}$~photon\,s$^{-1}$ 
using the basic photo-ionization equation from \citet{Osterbrock2006}, 
\begin{equation}\label{eqn:QHe}
\frac{{\rm Q(He}^+)}{{\rm photon~s}^{-1}} = 
      \frac{L({\rm He\,II}\lambda{4686})}{E_{\lambda{4686}}}\times
      \frac{\alpha_{\rm B}({\rm He}^+)}{\alpha_{\rm eff}({\rm He\,II}\lambda{4686})} \\
\end{equation}
\begin{equation}
         = 1.02\times10^{48}\frac{L({\rm He\,II}\lambda{4686})}{10^{36}~{\rm erg\,s}^{-1}},
\end{equation}
where $E_{\lambda{4686}}$, is the energy of the \heiiwr\ photon, and 
$\alpha_{\rm B}$ and $\alpha_{\rm eff}$ are case-B recombination coefficients.
We used the $\alpha_{\rm B}$ and $\alpha_{\rm eff}$ assuming reasonable
$T_{\rm e}$=10000~K and $n_{\rm e}$=100~cm$^{-3}$ for the nebula. The 
calculated Q(He$^+$) would change by less than 10\% for the entire range 
of $T_{\rm e}$ and $n_{\rm e}$ covered by the photoionized nebulae 
\citep{Kehrig2015}.

The total extinction-corrected \hb\ flux from the same regions is 
$4.44\times10^{-12}$~\ergcms. This is equivalent to a Lyman continuum rate 
Q(H$^0$) of $1.0\times10^{52}$~photon\,s$^{-1}$, using an equation similar 
to equation~\ref{eqn:QHe} but for Q(H$^0$), which is \\
\begin{equation}
\frac{{\rm Q(H}^0)}{{\rm photon~s}^{-1}} = 
         2.10\times10^{48}\frac{L(\rm H\beta)}{10^{36}~{\rm erg\,s}^{-1}},
\end{equation}
where $L(\rm H\beta)$ is the luminosity of the \hb\ line. The above equations 
result in the ratio 
\begin{equation}\label{eqn:qratio}
{\rm Q(He}^+)/{\rm Q(H}^0)=0.486\times\frac{L({\rm He\,II}\lambda{4686})}{L(\rm H\beta)}.
\end{equation}
The luminosity of \heii$\lambda{4686}$ corresponds to 2.05\% of the \hb\ luminosity,
which results in a value of Q(He$^+$)/Q(H$^0$)= 0.0100$\pm$0.0002.

The emission EW of \hb\ of a starburst population is a well-known 
age indicator. For unresolved regions, it is one of the quantities easiest to 
obtain from spectroscopic data. Furthermore, it is independent of extinction as long 
as the ionizing cluster and the nebula are spatially coincident, or there is 
no differential extinction between cluster stars and nebula. However, in a
resolved nebula such as the one discussed here, determination of EW(\hb) is
non-trivial. As the ionized nebula is spatially separated from the ionizing
cluster, we made the assumption that the photoionization from SSC-A is 
responsible for the total observed \hb\ flux. With this assumption, we divided
the observed integrated \hb\ flux by the continuum flux in the integrated 
spectrum of SSC-A (displayed in Fig.~\ref{fig:wr_candidates}), measured close 
to the \hb\ line. We obtained EW(\hb) under three assumptions. The first one 
is that the ratio is independent of extinction. In the second case, the 
continuum flux is corrected by \Av=2.3~mag, and the \hb\ flux is corrected for
the extinction determined by the Balmer decrement method. In the third case, the 
same \hb\ flux correction was used, but the continuum flux was corrected by
making use the minimum extinction of \Av=1.6~mag.
The three assumptions give values of 75~\AA, 75~\AA, 160~\AA, respectively.

\begin{figure*}
\begin{centering}
\includegraphics[trim=0cm 9.5cm 0cm 9.5cm, clip, width=0.80\linewidth]{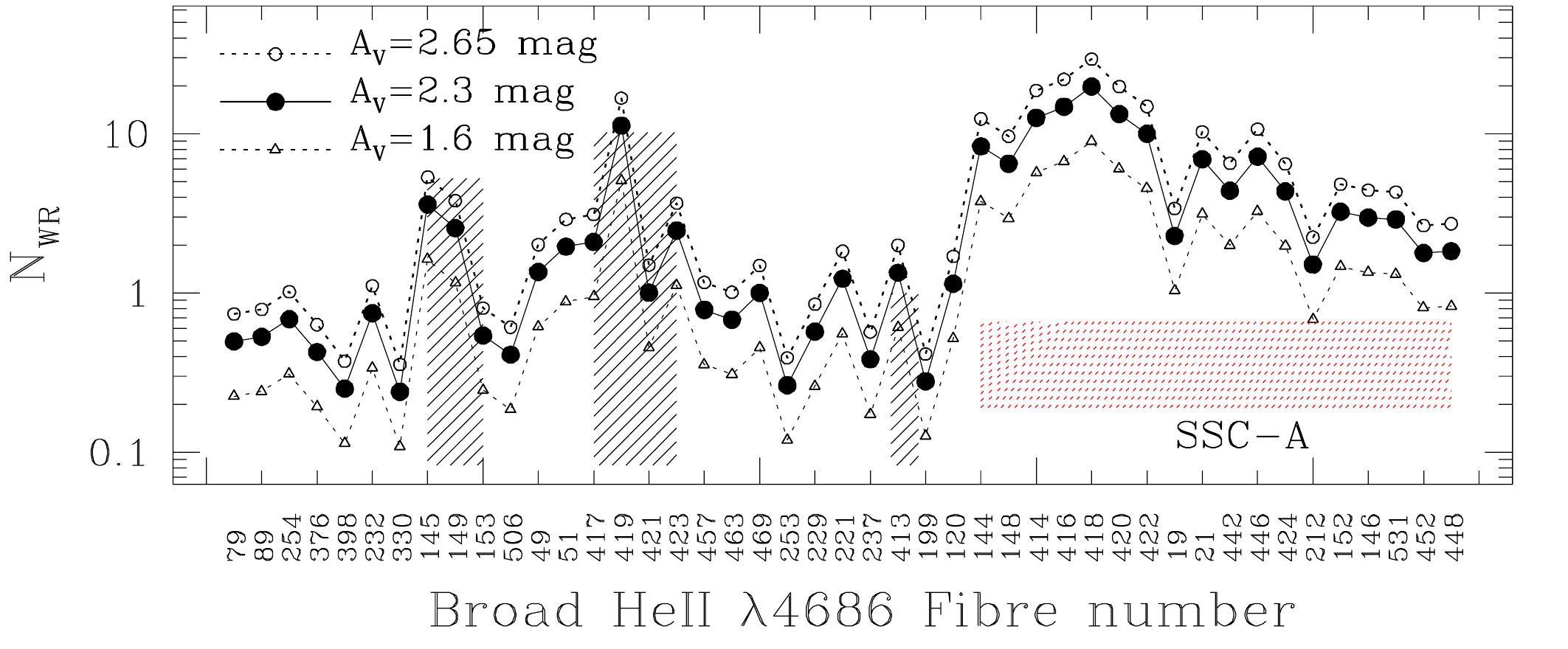}
\caption{
The number of WR stars, defined as the observed \heiiwr\ broad line luminosity
divided by the typical luminosity of a WNL star ($1.22\times10^{36}$~\ergs),
vs fibre number. All 50 individual fibres where we detected broad
\heiiwr\ feature are shown. The numbers are plotted for three values of 
extinction: \Av=1.6~mag, the minimum along the line of sight to NGC\,1569 
(triangles), \Av=2.65~mag, the mean determined for the ionized nebula (open 
circles), and \Av=2.3~mag, optimum value for SSC-A (solid circles).
Fibres belonging to SSC-A, and those that have cross-talk with fibres of SSC-A 
are indicated by red shaded area, and diagonally hatched area, respectively. 
SSC-A contains 124 WR stars for \Av=2.3~mag.
}
\label{fig:Nwr}
\end{centering}
\end{figure*}

\subsection{The number of WR stars in the mapped zone}

While fitting the BB with multi-Gaussians, we have looked for N (\niiiwr) 
and C (\ciiiwrb\ and \civwrb) broad lines in addition to the broad \heiiwr\ 
feature. In all spectra, including that of SSC-A, only one broad component was 
required, which is the \heiiwr\ line. In some spectra, a nebular \feiii\ line 
is seen at the expected broad \civwrb\ line, where our fitting procedure would 
have been able to recover a broad feature even in the presence of an overlying 
narrow line. The absence of any of the C lines points to the absence of WC 
stars in the observed zone. Previous observations of SSC-A that had covered the 
red bump (\civwrr) part of the spectrum had already indicated the absence of WC 
stars \citep{Rosa1997}, which is not unusual given that its metallicity 
is lower than solar \citep{Esteban2010}. 
Thus, we conclude that all our detections originated in WN 
stars. The relative weakness of \niiiwr\ suggests these are WNL stars.

We now estimate the number of WR stars in each of the fibres where the broad 
\heiiwr\ feature was detected in single-fibre spectra. As pointed out in \S3.2,
18 of these fibres belong to SSC-A, and 8 other fibres are in cross-talk with
fibres belonging to SSC-A. Single-fibre detections in rest of the fibre spectra 
are considered tentative, as the detected broad feature could not be recovered 
by summing spectra of neighbouring fibres. The number was estimated by dividing 
the observed luminosity in the \heiiwr\ broad feature by the typical luminosity 
of a WNL star which is $\sim$1.22$\times10^{36}$~\ergs\ \citep{Esteban2010}
for the metallicity of NGC\,1569. This value is $\sim$30\% smaller than the
corresponding value at the Solar metallicity \citep{Vacca1992}.
The observed luminosities of the \heiiwr\ broad feature
in each fibre need to be corrected 
for extinction. As discussed above, the \Av\ values derived at the positions of 
WR stars are highly unreliable, because of low nebular flux and high underlying 
Balmer absorptions at these locations. \citet{Larsen2011} found the colours of 
the resolved stellar population of SSC-A to be consistent with \Av=2.3 mag. This 
suggests that the internal extinction of the cluster stars is 0.35~mag lower 
than the mean value of the nebula, as expected for an attenuation law like the 
one proposed by \citet{Calzetti1994}. Hence, \Av=2.3 mag seems the most 
appropriate value to use to correct the fluxes of the \heiiwr\ broad feature 
from SSC-A. We hence calculate the number of WR stars using this optimum value 
\Av=2.3~mag. We also calculate the minimum and maximum number corresponding to 
\Av=1.6~mag and \Av=2.65~mag, respectively.

In Fig.~\ref{fig:Nwr}, we show the estimated number of WR stars in each fibre
where the broad \heiiwr\ feature is inferred, including those fibres with 
tentative detections. The solid symbols joined by solid lines indicate the number
using the optimum value for SSC-A of \Av=2.3~mag, whereas the other two lines 
correspond to minimum (\Av=1.6~mag, triangles joined by dotted lines) and mean 
nebular (\Av=2.65~mag, circles with dotted lines) extinction.
Fibres assigned to SSC-A are identified (red shaded area), as well the 8 fibres
that are affected by cross-talk with SSC-A (black diagonal hatched area).
With the optimum value of \Av=2.3~mag, all the fibres associated to SSC-A
have the luminosity of at least 1 WNL star, with one of the fibres (\#418) 
containing as many as 19 WNLs. This fibre belongs to the core of SSC-A. 
In total, we find 124$\pm11$ WNL stars in SSC-A by adding the numbers in each one
of the 18 fibres, where the quoted error assumes poissonian statistics.
This number agrees with the corresponding number obtained by fitting 
the multi-Gaussian on the spectra summed over 19 fibres covering the SSC-A.
Detected fluxes correspond to 56$\pm7$ and 186$\pm13$ WNL stars if we adopt
\Av=1.6~mag or \Av=2.65~mag, respectively. The uncertainty in the modeling of 
the extended wings of the spatial profile of the PSF of the fibres for those 
SSC-A spaxels where the WR bump is detected might lead to an underestimation 
to the total number of WR stars in SSC-A that could be of the order of a few 
per cent. Among the other single-fibre detections, only sources belonging to 
fibres \#49 and \#51 have $N_{WR}$ at least one for \Av$\ge$2.3~mag. Detected 
fluxes in the rest of the fibres are less than the flux expected for a WNL star.
As discussed in \S3.2, these detections were considered as tentative 
due to the non-recovery of the broad component in spectra obtained by 
summing spectra of neighbouring fibres. 

\citet{Rosa1997} have estimated the number of WR stars in SSC-A from their 
long-slit spectroscopic data assuming an \Av=1.6~mag. Using an older distance 
estimate of 2.2~kpc, 
and luminosity of a WNL star of $\sim$1.7$\times10^{36}$~\ergs\ from \citet{Vacca1992},
they estimated 24 WNL stars from their spectra, which corresponds to 66 WNL 
stars for the values used in our work. Our value for \Av=1.6~mag is comparable
with this value. On the other hand, the number of WR 
stars estimated based on the \hst\ narrow band image in the F469N filter by 
\citet{Buckalew2000} is 51$\pm$19 (\Av=1.6~mag, distance = 2.2~kpc) WNL stars, 
which corresponds to 141 WNLs for the distance and WNL luminosity used in our work.
This is 2.5 times higher 
than our value. As noted earlier (see Fig.~\ref{fig:wr_candidates} for SSC-A), 
nebular lines are weak for SSC-A, and hence it is unlikely that this discrepancy 
comes from the contribution of the nebular lines (e.g. \heiiwr, 
\hei$\,\lambda$4713\ and \feiii) to the inferred flux in the F469N filter.
Residual error in continuum subtraction because of the strong continuum of SSC-A 
could be the most likely reason for the overestimation of the number of WR stars.

\section{The ionization budget of \heii}

\subsection{Ionization by WR stars}

The region of NGC\,1569 analyzed in this work is dominated by the SSCs A and B 
in continuum light \citep{Arp1985}.  \citet{Prada1994} inferred the presence of 
Red Super Giants (RSGs) in both the SSCs A and B. SSC-A is made of two 
components, separated by 0.2~arcsec (3~pc), called A1 and A2 by 
\citet{DeMarchi1997}. \citet{Rosa1997} found spectral features characteristic 
of both the WR stars and RSGs in their ground-based long-slit spectra of SSC-A.
\citet{Origlia2001} analysed \hst\ UV spectra of SSC-A and suggested that the
RSGs and WR features are orginiated in A1 and A2 components, 
respectively, with the age of the latter component not exceeding 5~Myr.
\citet{Larsen2011} carried out photometry of the stars in the periphery of the 
clusters A and B on the \hst\ images to obtain turn-off ages in  the 
colour-magnitude diagram (CMD). They found RSGs in the CMD of SSC-B, but not 
in SSC-A. If the RSGs are restricted to the A1 component as \citet{Origlia2001}
suggested this implies this component is centrally concentrated and the stars 
in the periphery are part of A2, the younger of the two sub-clusters.
Based on the absence of RSGs in the CMD, they derived an upper age limit of 
4.5~Myr for this component. On the other hand, they obtained an age of 16~Myr 
for SSC-B. Thus, at present SSC-A is in the WR phase, when a cluster is most 
efficient in producing \heii\ ionizing photons. 
We point out that the bright \ha\ emission seen at the top-right corner of
the image presented in Fig.~\ref{fig:map_heiineb} is due to ionization by 
cluster 10 \citep{Westmoquette2007}, which coincides with WR cluster C1 of 
\cite{Buckalew2000}. We here investigate whether SSC-A alone is capable of 
producing the observed luminosity of the \heiiwr\ nebular line.

\begin{figure*}
\begin{centering}
\includegraphics[trim=1.0cm 3.2cm 8.0cm 2.0cm, clip, width=0.49\linewidth]{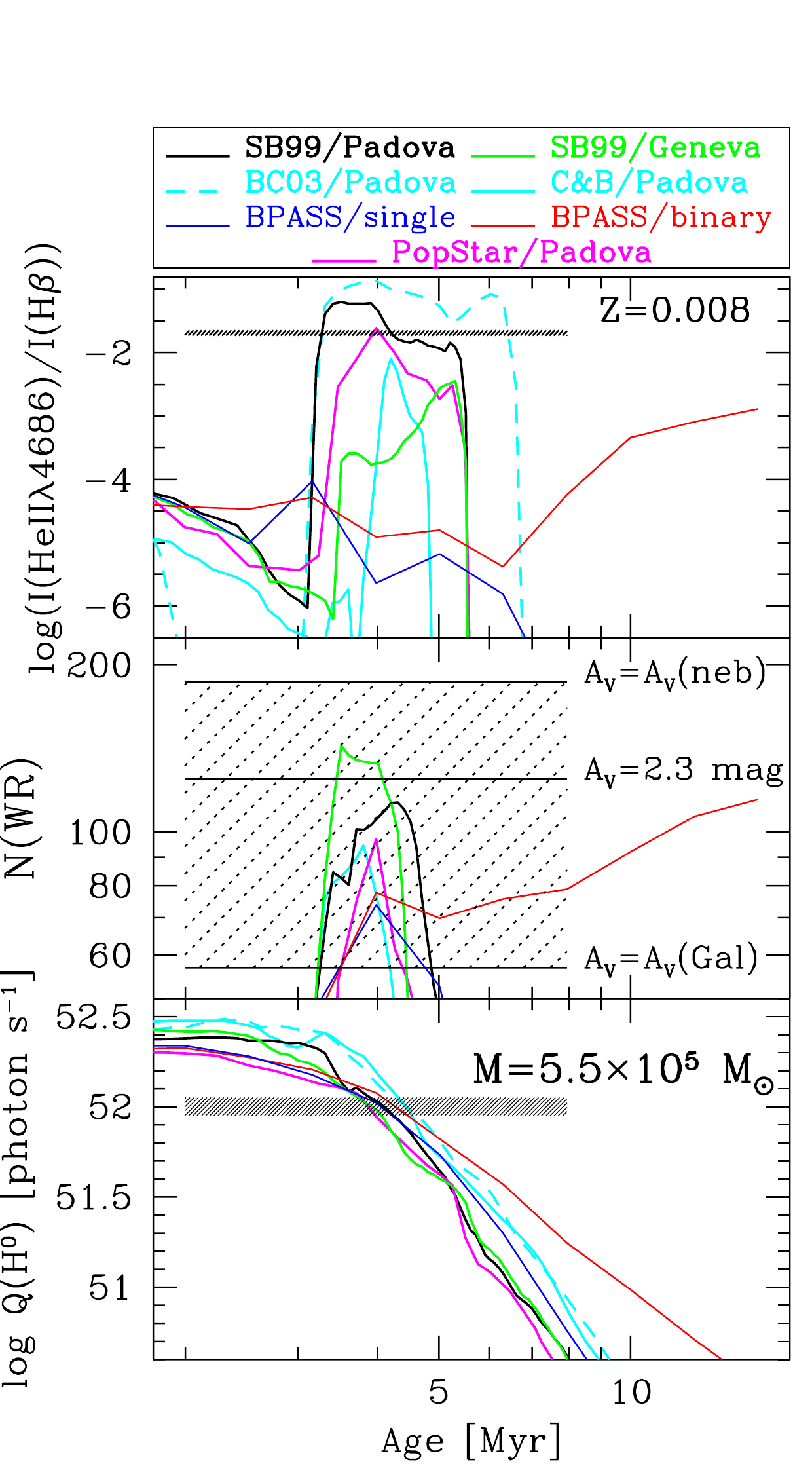}
\includegraphics[trim=1.0cm 3.2cm 8.0cm 2.0cm, clip, width=0.49\linewidth]{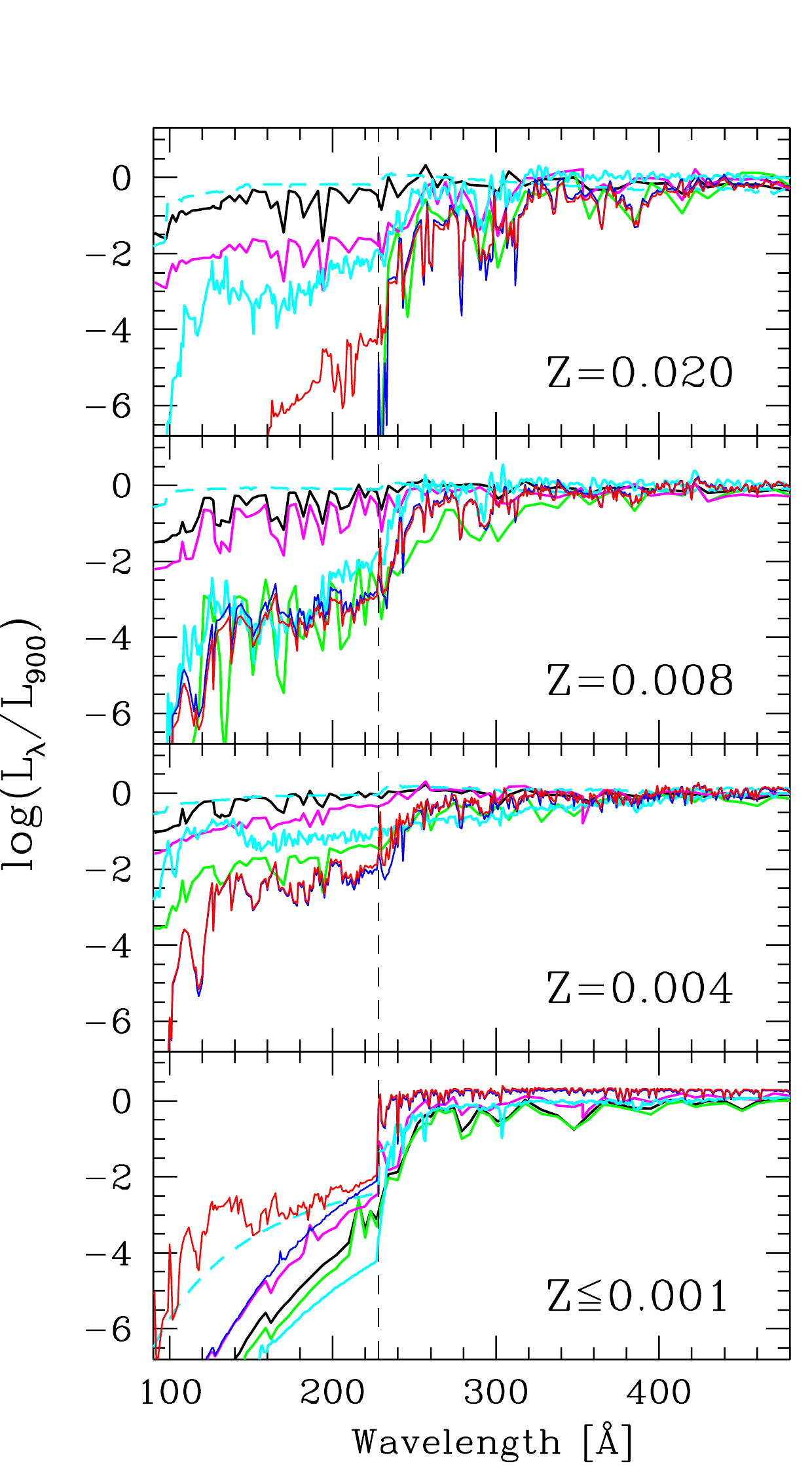}
\caption{
(Left panels) Comparison of observed quantities in NGC\,1569 with Population 
Synthesis models. The plotted observed quantities (horizontal hatched areas) 
are the ratio of integrated nebular \heiiwr\ line 
flux to that of \hb\ line (top-left), number of WR stars in SSC-A (middle-left) 
and the rate of hydrogen ionizing photons determined from the integrated \hb\ flux 
(bottom-left). Results from 7 different population synthesis models are shown, 
which are identified in a box above the plots (see text and Table~\ref{tab:ssps} for 
details of these models). Models that use Padova evolutionary tracks are able 
to reproduce the observed values for a cluster of mass=5.5$\times10^5$~\msol\ at 
an age of 4~Myr. All models use Kroupa IMF at Z=0.008, which is the metallicity 
closest to that of NGC\,1569 available in all the codes.
(right panels) Spectral shape near the He$^+$ ionizing edge (228~\AA) at the 4 
indicated metallicities during the WR phase. Spectra are 
shown normalised to their flux at 900~\AA\ wavelength. 
All spectra show a jump blueward of 228~\AA\ at the lowest plotted metallicity.
However, at higher metallicities the spectral shapes blueward of 228~\AA\ for
the plotted models do not coincide, with the spectra being harder in models 
incorporating Padova tracks as compared to those obtained using other tracks.
}
\label{fig:starburst99}
\end{centering}
\end{figure*}

The principal sources of \heii\ ionizing photons in young stellar systems are 
O stars and their evolutionary products, such as WR stars.
These stars have extended atmospheres with high mass-loss rates, hence a
calculation of the rate of \heii\ ionizing photons has to take into account 
the radiation transfer through these atmospheres. In recent 
years, such spectra are available from {\sc CMFGEN} code \citep{Pauldrach2001, 
Hillier1998, Smith2002} for O and WR stars, and from the Potsdam {\sc PoWR} library 
\citep{Grafener2002} for WR stars. 
We compared our observed values with publicly available Simple Stellar 
Population (SSP) models that have incorporated either CMFGEN or PoWR spectra 
in their codes. Calculations are available at discrete values of metallicity. 
The gas-phase oxygen abundance of NGC\,1569 corresponds to Z=0.006 for a depletion
of 30\% of oxygen on to dust grains and assuming [O/Fe]=0.0 \citep{Gutkin2016}.
We illustrate the results for models using Z=0.008, which is the closest 
metallicity to that of NGC\,1569, and comment on the metallicity dependence of 
the obtained results. A summary of the Z=0.008 models we used is given in 
Table~\ref{tab:ssps}.
\begin{table*}  
\caption{Comparison of Population Synthesis model results for Z=0.008 with observations of SSC-A}\label{tab:ssps}
\begin{tabular}{llccccccl}
\hline
SSP code ID & WR model + atmosphere & \multicolumn{2}{c}{I(\heii)/\hb} & \multicolumn{2}{c}{WR phase} & EW(\hb) & Q(H0) & Comments \\
            &                       & age & peak     & age  &  $N_{\rm WR}$ &        &    log      & \\
            &                       & Myr & log      & Myr  &               &  \AA\  & ph s$^{-1}$ & \\
    (1)     &      (2)              & (3)    &   (4)        & (5)     &    (6)       & (7) & (8) & (9) \\
\hline
SB99/Padova    & Padova1994+CMFGEN & 3.5 & $-$1.20 & 3.3--4.9 & 123 & 278--88 & 52.03 & Good fit \\ 
SB99/Geneva    & Geneva1994+CMFGEN & 5.3 & $-$2.45 & 3.2--4.4 & 155 & 157--85 & 51.98 & I(\heii)/I(\hb) too low\\
PopStar/Padova & Padova1994+CMFGEN & 4.0 & $-$1.61 & 3.5--4.5 & 106 & 165--84 & 51.95 & Good fit\\
BC03/Padova  & Padova1994+PoWR   & 4.0 & $-$1.03 &  ---     & --- & ---     & 52.13 & Good fit\\
C\&B/Padova  & Padova2015+PoWR   & 4.2 & $-$2.10 &  3.3--4.2&  89 & ---     & 52.10 & Marginal fit\\
BPASS/single   & Cambridge+PoWR    & 3.2 & $-$4.03 & 4.0--5.0 &  80 & 180--110& 52.02 & I(\heii)/I(\hb) too low \\
BPASS/binary   & Cambridge+PoWR    & 16  & $-$2.89 & 4.0--16  & 124 & 195--19 & 52.08 & I(\heii)/I(\hb) too low\\
\hline
SSC-A          & Observed or inferred   & & $-$1.69 & 4.0$\pm0.5$ &  124 & 75  & 52.00 & $(5.5\pm0.5)\times10^5$~\msol\ \\
               &                        & &         &             & 56--186    &  160  &  &  \\
\hline
\end{tabular}
(1)--(2): Model name, evolutionary tracks and WR atmospheric models used (Padova1994=\citealt{Padova1994};
Geneva1994=\citealt{Geneva1994}; Padova2015=\citealt{Chen2015}).
Last row contains observationally inferred quantities. 
The tracks used in PopStar is a modified version of Padova1994 tracks as described in \citet{PopStar2009};
(3) age at which nebular I(\heiiwr)/I(\hb) ratio is maximum; 
(4) maximum value of I(\heiiwr)/I(\hb) in the model in log units;
(5) The age interval over which the model 
has more than the observationally estimated minimum number of 56 WR stars, for the determined mass. 
The last row contains the age range over which models using Padova tracks reproduce the observed I(\heiiwr)/I(\hb);
(6) The maximum number of WR stars in the model in the WR phase. 
Last row contains the best estimation and the possible range depending on the assumption on extinction;
(7) range of EW(\hb) for the age range in column 5. EW(\hb) decreases monotonically with age. 
Observed range in last row corresponds to the minimum and maximum values based on different assumptions on
differential extinction (see text for details);
(8) model log(Q(H$^0$)) at the most likely age (4 Myr) and mass (5.5$\times10^5$~\msol);
(9) comments on the comparison between the observations and model. Last row contains the best 
determined mass and error on it, which takes into account the error on the determined age.
\end{table*}

\begin{figure}
\begin{centering}
\includegraphics[trim=0.5cm 3.5cm 1cm 4.0cm, clip, width=0.98\linewidth]{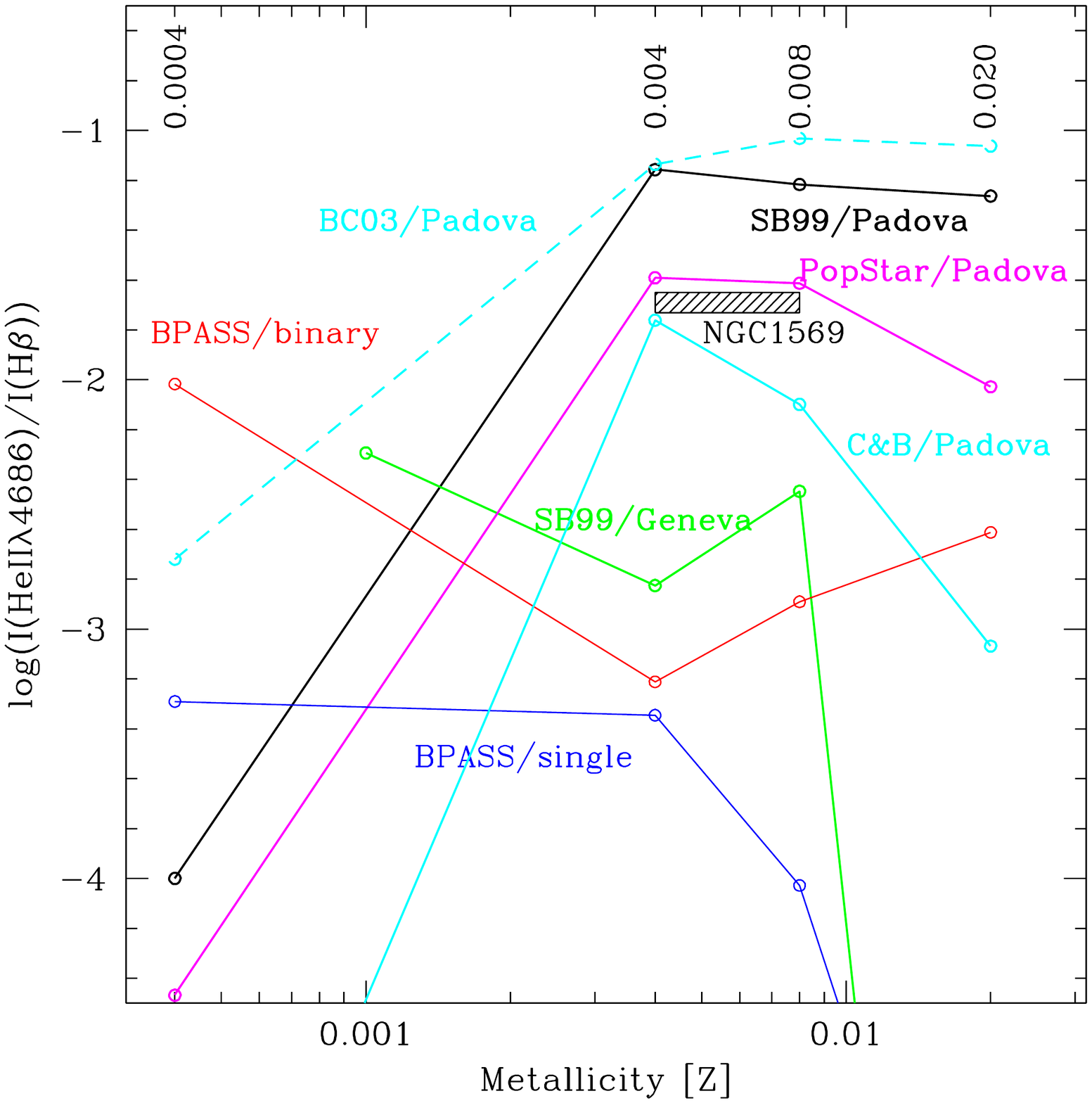}
\caption{
Maximum values of the nebular I(\heiiwr)/I(\hb) ratio (points joined by lines) 
reached as a function of the abundance for the SSP models indicated close to 
each curve. The observed value for NGC\,1569 is indicated, where the range
in metallicity takes into account 10--30\% depletion of oxygen into dust
grains \citep{Gutkin2016}.
The observed values are in agreement with the
highest values of model I(\heiiwr)/I(\hb) ratio, that happens between Z=0.004 and
Z=0.008 for new (C\&B/Padova) as well as old Padova (SB99/Padova, PopStar/Padova) models.
}
\label{fig:fqz}
\end{centering}
\end{figure}

The codes we used are, STARBURST99 \citep{Starburst99, Starburst2014} with 
Padova \citep[SB99/Padova;][]{Padova1994} and 
Geneva high-mass loss \citep[SB99/Geneva;][]{Geneva1994} evolutionary tracks, 
PopStar \citep[PopStar/Padova;][]{PopStar2009} with an updated version of the 
\citet{Padova1994} Padova tracks, two versions of GALAXEV: BC03/Padova 
\citep{Bruzual2003} based on tracks from \citet{Padova1994} and C\&B/Padova 
\citep[Charlot \& Bruzual, in preparation; see][]{Gutkin2016, Plat2019}
based on tracks from \citet{Chen2015}, and the BPASS \citep{Eldridge2017} 
that uses Cambridge evolutionary tracks in single (BPASS/single) and binary (BPASS/binary) mode. 
STARBURST99 and PopStar make use of CMFGEN code, whereas GALAXEV and BPASS use PoWR code,
to model the extended atmospheres of WR stars. We downloaded the latest results from these 
codes from the websites of the respective codes, uniformly using 
\citet{Kroupa2001} IMF between 0.15~\msol\ to 100~\msol. In the case of BPASS 
models, the detailed calculations of ratios of different lines, including that 
of I(\heiiwr)/I(\hb), has been provided by \citet{Xiao2018}.
We used the ratios corresponding to the ionization parameter of $\log U=-$1.5 
and an atomic density of 1~atom\,cm$^{-3}$. For the rest of the models,
we used Eqn.~\ref{eqn:qratio} to convert the ratio of He$^+$ to H$^0$ ionizing 
photon rates to a flux ratio I(\heiiwr)/I(\hb).

Results of the comparisons are presented in Fig.~\ref{fig:starburst99} on the
left panels. Horizontal hatched areas in each panel encompass the entire range
of observable quantities with account for observational errors.
For I(\heiiwr)/I(\hb) and Q(H$^0$), errors are taken as 10\%, whereas for the 
number of WR stars, the main source of error is the uncertainty in extinction correction.

The model Q(H$^0$) decreases monotonically with age, whereas the 
nebular I(\heiiwr)/I(\hb) ratio has a well-defined peak at $\sim$4~Myr, which 
corresponds to the appearance of WR stars in the cluster. 
This ratio is independent of distance, cluster mass, extinction, the 
chosen nebular parameters (temperature and density), and the 
choice of IMF parameters, as long as the upper cut-off mass is not very much
different from 100~\msol. Hence, among the three plotted quantities, the age 
derived from I(\heiiwr)/I(\hb) ratio is the most reliable. 
Three models that use \citet{Padova1994} Padova evolutionary tracks have peak 
values of I(\heiiwr)/I(\hb) higher than the observed value in NGC\,1569 (see 
Tab.~\ref{tab:ssps}). The ratio is marginally smaller for model that makes use of
the results from the new Padova tracks \citep{Chen2015}, whereas in the other 
three models this ratio is much lower.
All the models reproduce the observationally estimated Q(H$^0$) for a cluster 
of mass (5.5$\pm0.5$)$\times10^5$~\msol\ in its WR phase. 
The observed EW(\hb) is also consistent with the value expected during the
WR phase for single star evolutionary models (see column~7 in Tab.~\ref{tab:ssps}).
The number of WR stars predicted in all models agrees with the best estimation 
of the observed value to within 30\% for the mass inferred above.
The WR phase in single star evolutionary models lasts between $\sim$3.5--5~Myr. 
The limits for star cluster age are determined from the condition that the 
model predicted number of WR stars for the above inferred mass exceeds the 
lower limit for the observationally inferred number of WR stars in SSC-A. 
This age range, as well as the age at which I(\heiiwr)/I(\hb) ratio reaches its peak 
value, are given in Table~\ref{tab:ssps}, in columns 5 and 3, respectively.

It is interesting to note that the SB99/Padova and SB99/Geneva models use the 
same atmospheric models in the WR phase, but Geneva evolutionary tracks 
have lower I(\heiiwr)/I(\hb) in spite of having $\sim$25\% more number 
of WR stars. This implies that the low values of I(\heiiwr)/I(\hb) in 
SB99/Geneva models are not due to lack of WR stars, but due to the different 
surface parameter values in these models as compared to those in the Padova tracks. 
Both the single and binary BPASS models predict systematically a smaller number of WR 
stars, and a lower peak value of I(\heiiwr)/I(\hb) at the 3--5~Myr age. 
In binary models, peak values of these two parameters are reached at later 
ages ($\sim$16~Myr) when Q(H$^0$) falls by a factor of 30 with respect to 
its values at the WR phase in single star evolutionary models. The observed 
EW of \hb\ is not in favour of this advanced age.

In the right panel of Fig.~\ref{fig:starburst99}, we plot the typical emergent 
spectum around the He$^+$ ionization edge (228~\AA; the dashed vertical line)
during the WR phase (4~Myr for all models except BPASS where it is 3.2~Myr)
for a range of metallicities. All spectra are normalized to their fluxes at $\lambda$=900~\AA.
The normalisation wavelength is specifically chosen to be slightly blueward of the H$^0$ 
ionization edge (912~\AA), so that the shape of the plotted spectra blueward 
of 228~\AA\ is an indicator of the I(\heiiwr)/I(\hb) ratio.
The spectra in the four models incorporating 
the Padova tracks (SB99/Padova, PopStar, BC03 and C\&B) exhibit more He$^+$ 
ionizing photons blueward of 228~\AA\ for metallicities Z$\geq$0.004 than
those in the SB99/Geneva and BPASS models. At these metallicities, the emergent 
spectrum using the new Padova tracks (C\&B/Padova) is softer than that using the 
older Padova tracks, but it is still harder than that obtained in 
SB99/Geneva and BPASS models. On the other hand, the spectral behaviour
is similar in all the seven plotted models at low metallicities (Z$\leq$0.001),
all models predicting a downward jump at 228~\AA. 
The plot illustrates that the 
atmospheric parameters (effective temperature and bolometric luminosity) of WR 
stars in Padova tracks are particularly able to account for the observed 
ratio of I(\heiiwr)/I(\hb) at metallicities Z$\geq$0.004.

In the above analysis we have compared the values of NGC\,1569 with Z=0.008 
metallicity models. For the observed gas-phase abundance of oxygen, the 
metallicity Z can be as low as Z=0.004 if less than 10\% of oxygen is depleted 
on to dust grains \citep{Gutkin2016}. Given that the hardness of the spectrum 
blueward of the He$^+$ ionization edge depends on metallicity, we now examine 
the behaviour of I(\heiiwr)/I(\hb) ratio with metallicity during the WR phase. 
This is illustrated in Fig.~\ref{fig:fqz} for all the models discussed in this work. 
The observed ratio in NGC\,1569 is well reproduced for a metallicity range 
of Z=0.004--0.008 in all SSPs that use Padova tracks, independent of whether 
CMFGEN or PoWR models are used to represent the atmospheres of WR stars. 
In comparison, SSP models using Geneva tracks and BPASS models predict
more than an order of magnitude lower values. 

It can be seen in Fig.~\ref{fig:fqz} that the I(\heiiwr)/I(\hb) ratio 
decreases sharply at Z$<$0.004 in models using Padova tracks. 
The maximum I(\heiiwr)/I(\hb) ratio at the lowest plotted metallicity (Z=0.0004)
corresponds to the BPASS/binary model. However, the ratio just about reaches 0.01.
The observed ratio
in metal-poor galaxies is often higher that these predicted values \citep[see][]{Shirazi2012}.
In fact, observed data show a gradual tendency for the nebular I(\heiiwr)/I(\hb)
ratio to increase with decreasing metallicity reaching values as high as $\sim$0.06
in the most metal-poor galaxies known \citep{Schaerer2019}.
This incapability of the SSP models to reproduce the observed I(\heiiwr)/I(\hb) ratio
at low-metallicities (Z$<$0.004) is often referred to as the He$^+$ ionization budget problem.
The problem starts arising just below the metallicity of NGC\,1569.
At these low metallicities, the problem is aggravated due to the small number of 
WR detections.
For example, I\,Zw\,18, one of the most metal-poor galaxies, emits as much He$^+$ ionizing photons as
SSC-A in NGC\,1569, but it has at the most 9 WR stars detected \citep{Kehrig2015}
as compared to the 124 WR stars in SSC-A. 
A comprehensive analysis of this problem was carried out recently by \citet{Plat2019}.
They used the same C\&B/Padova model that we have used here, and calculated the I(\heiiwr)/I(\hb)
ratio for a variety of additional input physics not explored in this study.
They find that the conditions most favourable to produce I(\heiiwr)/I(\hb)$>0.01$ 
at low metallicities include:
(1) the presence of stars significantly more massive than 100~\msol, 
(2) extremely high ionization parameter, $\log(U)>-1$, 
(3) the presence of interacting binaries that produce X-rays,
(4) ionization of He$^+$ by radiative shocks, or 
(5) when analysing intergrated spectra of distant galaxies, ionization of He$^+$ by an active galactic nucleus. 
It is likely that more than one of these conditions are met in some of the metal-poor galaxies. 

\begin{figure*}
\begin{centering}
\includegraphics[width=0.80\linewidth]{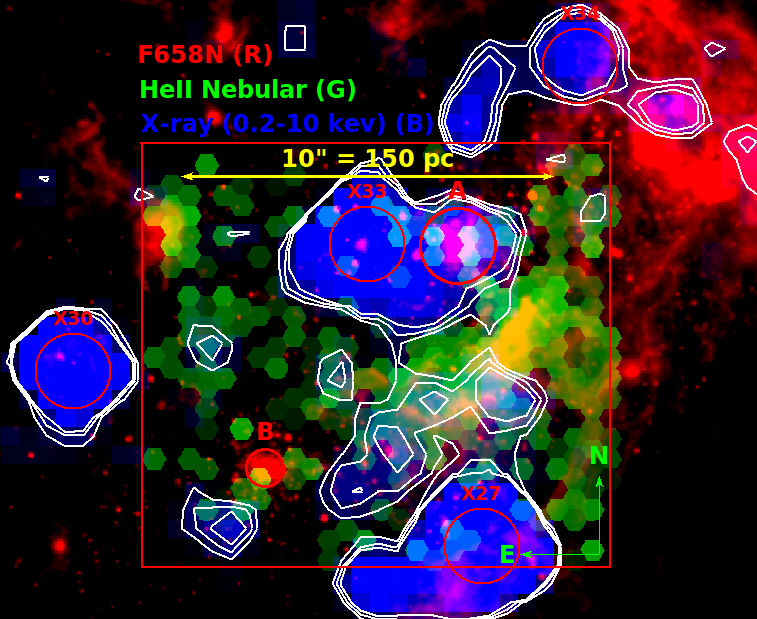}
\caption{
X-ray emission map from the Chandra/ACIS image (0.2--10~keV band)
from \citet{Monica2015} in blue is superposed on the \heii\ nebular 
map in green and F658N (\ha+continuum) in red. The contours show 
the X-ray emission at 3 levels (3$\sigma$, 3.5$\sigma$ and 4$\sigma$
above the background) to help see the position of the faint X-ray 
diffuse emission with respect to the \heiiwr\ nebular emission.
Most of the X-ray emission within the MEGARA FoV (red square) originates in two 
hard X-ray point sources (X27 and X33), which are marked by red circles. 
The diffuse X-ray emission coincides with part of the \heiiwr\ nebular emission.
However, there is no X-ray emission at the brightest zone of the nebular 
\heiiwr\ emission.
}
\label{fig:map_xray}
\end{centering}
\end{figure*}

In summary, the observed number of Q(H$^0$) from nebula surrounding SSC-A 
and $N_{\rm WR}$ in SSC-A are consistent with each other 
for all models for a cluster mass of (5.5$\pm0.5$)$\times10^5$~\msol\ at 
$\sim$4.0$\pm0.5$~Myr age. The observed nebular I(\heiiwr)/I(\hb) ratio is in the range of 
predicted values during the WR phase in models that use Padova evolutionary tracks.
The inferred age for SSC-A is in agreement with the age of 4.5~Myr 
determined by \citet{Larsen2011} using the CMD of stars in the periphery of SSC-A.
The inferred mass is $\sim$15\% lower as compared to the photometric 
mass of $6.3\times10^5$~\msol\ determined by \citet{Larsen2011}, 
after scaling their mass to the distance of 3.1~Mpc used in this work. 
Slightly larger mass for SSC-A in \citet{Larsen2011} is expected given that
the SSC-A has two populations, a centrally concentrated population (A1) 
containing RSGs (older than 7~Myr) and a slightly extended component (A2) 
containing WR and O stars. Our measured mass is based on the ionizing flux 
and hence corresponds to the mass of the component A2. On the other hand, the 
photometric mass derived by \citet{Larsen2011} is based on integrated photometry 
of stellar light, and hence it includes the mass of both the components. 
On the other hand, the mass derived by us is $\sim$25\% higher than
the dynamical mass of $4.1\times10^5$~\msol\ obtained by \citet{Ho1996} using
spectral lines originating in cool supergiants. \citet{Ho1996} commented that 
the velocity dispersion obtained from cool supergiants in the integrated spectrum 
could underestimate the mass by as much as a factor of 2. 
Furthermore, the cool supergiants belong to the older of the two populations.
The derived dynamical mass is expected to be the total of the two populations 
if they are dynamically mixed. The presence of hot massive stars, but not cool 
supergiants, in the periphery of SSC-A argues against such a mixing
\citep{Larsen2011}. 
Measurements of velocity dispersion using spectral lines sensitive to 
hot stars (e.g. He absorption lines which are prominent in the
spectrum of SSC-A) could help to address this issue. 
MEGARA has the capability of obtaining velocity dispersion in its
high resolution mode, providing a possibility of addressing this issue in the near future.

\subsection{Morphology of the \heiiwr\  nebula and the location of SSC-A}

We found that the entire observed \heiiwr\ emission can be understood in terms 
of the ionization from WR stars in SSC-A. However, the observed \heiiwr\ 
emission is not coincident with the location of the ionizing cluster. Instead, 
the most intense part of the ionized nebula both in the Balmer  and  in the 
\heiiwr\ lines lies $\sim$40~pc to the south-west of the cluster. 
The zone of intense emission is part of a semi-circular arc of 150~pc
diameter, with its centre $\sim$40~pc to the 
east of SSC-A. Because of this off-centring, the distance of ionizing source 
to different parts of the emitting arc is different, causing the surface 
brightness of the emission to decrease along the arc as its distance increases 
from SSC-A. The emission is weak inside the semi-circular arc
(see Section~3). The observed 
morphology resembles the structure of a classical wind-driven bubble. If this 
is the case, the hot shocked gas that fills the bubble should be a source of 
the X-ray emission \citep{Weaver1977, Chu1990, Silich2005}. Soft X-rays are 
indeed detected in NGC\,1569, whose morphology is discussed below.

\subsection{X-ray morphology}

The combined effect of stellar winds from massive stars in cluster A,
including those from the multiple WR stars, is to form a global star 
cluster wind able to expel the inserted and the residual gas away into 
the interstellar medium (ISM). The interaction of the combined star 
cluster wind with the ISM leads to a strong shock which sweeps up the 
ambient gas into a thin shell, while the cluster wind is thermalised 
at a reverse shock. The shell of swept-up matter cools rapidly, if 
the ambient gas density is not too small, to be then completely
or partially photoionized by the energetic photons escaping
the cluster \citep{Castor1975, Martinez-Gonzalez2014}.

Both soft (0.2--2~keV) and hard (2--10~keV) X-ray emission have been detected 
from the central zone of NGC\,1569 \citep{Martin2002, Monica2015}. The soft 
X-ray emission is diffuse, whereas the hard X-rays principally come from point 
sources. We plot the X-ray map (blue) superposed on the \heii\ nebular map 
(green) in Fig.~\ref{fig:map_xray}. The F658N image (\ha\ + continuum 
emission) is shown in red for positional reference. The observed FoV contains 
two hard X-ray emitting point sources catalogued by \citet{Monica2015}:
source 27 (identified as X27) is associated to an X-ray binary, and source 33 
(denoted as X33) as originating in the cluster 29 of \citet{Hunter2000} (see 
our Fig.~\ref{fig:wrimages}).

Most of the soft X-ray emission comes from diffuse regions coincident with 
parts of the \heii\ nebula. However, the brightest part of the \heii\ nebula
does not show any X-ray emission.  As illustrated by \citet{Monica2015}, 
there is faint emission from the intervening zone between the SSC-A and the 
ionized nebula. This emission is likely associated with the hot shocked winds.

X-ray photons can also produce the \heii\ emission. 
We used the factor $q={\rm Q(He}^+)/L_X=2\times10^{10}$ photon\,erg$^{-1}$ 
defined by \citet{Schaerer2019} to estimate the Q(He$^+$) using the observed 
X-ray luminosity from the two X-ray sources in our FoV. The two sources combined
contribute $\sim1.7\times10^{48}$~photon\,s$^{-1}$, which is less than 2\% of 
the observed ionization requirement. Hence, we are justified 
in neglecting any contribution to ionization from X-ray photons.

\section{Conclusions}

Using the recently available integral field spectrograph MEGARA at the 10.4-m 
GTC, we detect extended \heiiwr\ nebula in the central starburst zone of the 
nearby dwarf galaxy NGC\,1569. The nebula extends along a semi-circular arc
of $\sim$150~pc (10~arcsec) diameter and $\sim$40~pc width, with the zone of 
the brightest \heiiwr\ nebular emission lying at $\sim$40~pc (2.5~arcsec) 
south-west of the massive young cluster SSC-A. The spectral data also show 
broad \heiiwr\ emission from 18 individual fibres belonging to the SSC-A. 
None of the other sources that are well separated from SSC-A, where WR 
detection have been reported using \hst\ imaging data by \citet{Buckalew2000}, 
show broad \heiiwr\ emission. We find that these sources are stars and clusters 
immersed in the \heiiwr\ nebula. We use the \hb\ and \hg\ nebular lines in the 
same spectra to map the extinction in the observed zone. The minimum extinction 
we obtain is consistent with the \Av=1.6~mag from the Milky Way along the line 
of sight to NGC\,1569. The mean value for all fibres where we could reliably 
measure \Av\ is \Av=2.65~mag. The low surface brightness of the nebular flux 
in and around SSC-A prevented us from obtaining nebular extinction towards 
this cluster. Using \Av=2.3~mag, the value used by \citet{Larsen2011} to 
interpret the colours of the resolved stellar population in SSC-A, we estimate 
124$\pm11$ WR stars of type WNL in SSC-A. We derive hydrogen and He$^+$ 
ionizing rates of 1$\times10^{52}$~photon\,s$^{-1}$ and 1$\times10^{50}$~photon\,s$^{-1}$,
respectively. These observed quantities 
are in agreement with the expectations from single stellar population models 
at 4.0$\pm$0.5~Myr for a cluster of mass (5.5$\pm0.5$)$\times10^5$~\msol.
The derived age is consistent with the turn-off ages determined by 
\citet{Larsen2011} from resolved stellar populations in the periphery of this cluster. 
A careful comparison of the most commonly used population synthesis models, 
we find that the predicted values of nebular I(\heiiwr)/I(\hb) in
models that make use of Padova evolutionary tracks are in the range of
observed value of 0.02 for metallicities Z$\ge$0.004.
Thus, at $12+\log({\rm O/H})$=8.19 (Z$\sim$0.006), which is the measured 
value for NGC\,1569, the WR stars from the cluster are able to explain the 
origin of the \heiiwr\ nebula.

\section*{Acknowledgements}

It is a pleasure to thank the anonymous referee for thoughtful comments
that lead to a significant improvement of the manuscript.
This publication is based on data obtained with the MEGARA instrument at the 
Gran Telescopio Canarias, installed in the Spanish Observatorio del Roque de 
los Muchachos, in the island of La Palma. MEGARA has been built by a Consortium led by the 
Universidad Complutense de Madrid (Spain) and that also includes the Instituto 
de Astrof\'{i}sica, \'Optica y Electr\'onica (Mexico), Instituto de Astrof\'{i}sica 
de Andaluc\'{i}a (CSIC, Spain) and the Universidad Polit\'ecnica de Madrid (Spain). 
MEGARA is funded by the Consortium institutions, GRANTECAN S.A. and European 
Regional Development Funds (ERDF), through Programa Operativo Canarias FEDER 2014-2020.
We thank CONACyT for the research grants CB-A1-S-25070 (YDM), CB-A1-S-22784 (DRG), and 
CB-A1-S-28458 (SS). VMAGG is funded by UNAM DGAPA PAPIIT project number IA100318.
Authors acknowledge financial support from the Spanish MINECO under grant
numbers AYA2016-79724-C4-4-P, AYA2016-75808-R, AYA2017-90589-REOT and RTI2018-096188-B-I00.

\section{Data availability}

The fluxes of principal emission lines used in this work are available 
in the article and in its online supplementary material. The reduced fits files
on which these data are based will be shared on reasonable request to the first 
author.

\appendix

\section{Measured data in individual fibre spectrum} \label{apen:data}

\begin{figure*}
\begin{centering}
\includegraphics[width=0.9\linewidth]{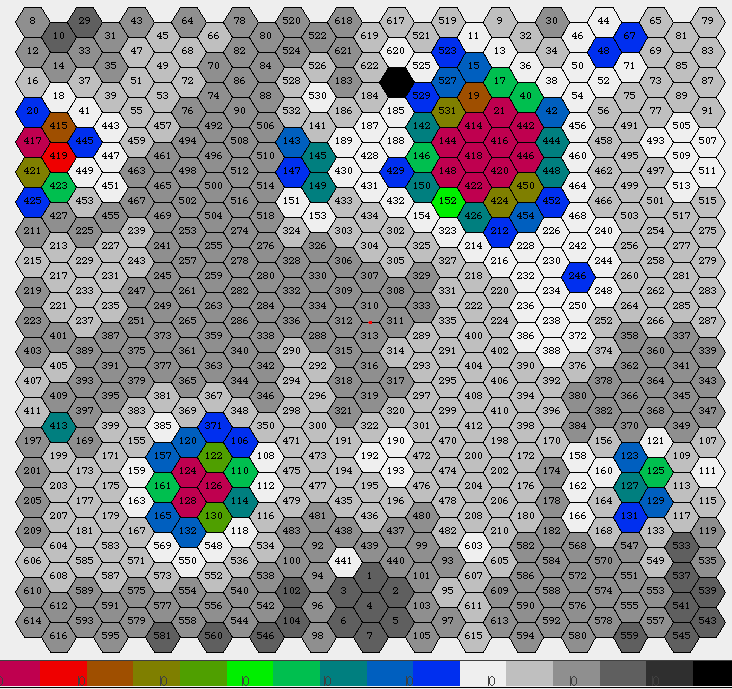}
\caption{
Spaxel (hexagon) map of the region of NGC\,1569 observed by MEGARA. The spaxel contains
the summed flux from the entire spectra. The colour-scale at the bottom: 
the black colour indicates faint regions and the red the brightest regions. 
The numbers indicate the fibre identification numbers. The fibres nearest to 
the centres of SSC-A and SSC-B are \#418 and 128, respectively. All coloured 
hexagons in the same rows but to the left of SSC-A, and to the right of SSC-B 
are artefacts due to the cross-talk with the fibres pointed to these two bright 
clusters, respectively.
}
\label{fig:map_qla}
\end{centering}
\end{figure*}

In Table~\ref{table}, we give the measured and calculated values for all the 
lines used in this work for 50 illustrative fibre spectra. The first 25 rows
contain data for fibres arranged as a decreasing function of \hb\ flux
(column 6), whereas the remaining 25 rows show data for fibres where we 
detected WR features. These data are arranged in the decreasing order of 
$N_{\rm WR}$ (column 19). The electronic version contains data for all the 567 
fibres, arranged in the increasing order of fibre number (column 1). The note 
to the table contains a detailed explanation of the quantities in each column.
In Fig.~\ref{fig:map_qla}, we show a spaxel map of MEGARA, where we indicate
the location of each fibre, identified by its number.

\begin{table*}  
\small\addtolength{\tabcolsep}{-2pt}
\caption{\heiiwr\ nebular and WR parameters measured in individual fibre spectra}\label{table}
\centering
\begin{threeparttable}
  \begin{tiny}
\begin{tabular}{@{}rrrrrrrrrrrrrrrrrrrrll@{}}
\hline
\multicolumn{5}{c}{Fibre coordinates} &\multicolumn{4}{c}{\hb-related}&\multicolumn{2}{c}{\hg-related}& \multicolumn{5}{c}{He{\sc ii}$\lambda$4686 nebula-related}    &    \multicolumn{4}{c}{He{\sc ii}$\lambda$4686 WR-related}& \multicolumn{2}{c}{Extinction} \\
Fno& dX & dY   & R   & PA  &$f_T$& snr & EW& FW&f$\beta$& snr&$\lambda_0$ & FW& $f_T$ &  snr & f$\beta$ & $\lambda_0$& FW&$N_{\rm WR}$ & e$N$  &\Av  & e\Av  \\
   & arcsec & arcsec & pc   & $^\circ$  & &  & \AA\ & \AA\ &  & &\AA\ & \AA\ &  &   &  & \AA\ & \AA\ & &  & mag  & mag  \\
  1&   2&     3&     4&    5&     6&      7&    8&    9&    10&  11&     12&   13&  14&      15&  16&    17&    18&   19&   20&  21&   22\\
\hline
242 & 1.86 & -1.61 &  36.9 & 229.1 & 6.181 & 82.7 & 51.2 &  1.2 & 19.6 &10.8 & 4685.1 &  4.1 &10.230 &   8.8 & 3.7  &   0.0  & 0.0 &  0.0 &  0.0 & 4.52 & 0.43\\
240 & 2.32 & -1.34 &  40.3 & 240.0 & 3.990 & 56.2 & 41.1 &  1.3 & 19.6 & 9.1 & 4684.7 &  2.5 & 8.726 &  10.7 & 4.9  &   0.0  & 0.0 &  0.0 &  0.0 & 4.40 & 0.48\\
244 & 2.32 & -1.88 &  44.9 & 231.0 & 3.972 & 59.9 & 57.1 &  1.2 & 21.8 & 9.6 & 4684.9 &  2.9 & 6.808 &   9.6 & 3.8  &   0.0  & 0.0 &  0.0 &  0.0 & 4.12 & 0.48\\
246 & 1.86 & -2.15 &  42.7 & 220.8 & 3.010 & 70.5 & 64.5 &  1.2 & 23.4 &16.7 & 4684.8 &  2.1 & 4.168 &  10.3 & 3.1  &   0.0  & 0.0 &  0.0 &  0.0 & 3.65 & 0.25\\
406 & 0.46 & -4.03 &  60.9 & 186.6 & 2.171 & 67.0 & 50.5 &  1.3 & 23.0 &14.1 & 4685.1 &  2.1 & 2.881 &   9.2 & 2.9  &   0.0  & 0.0 &  0.0 &  0.0 & 3.71 & 0.30\\
 20 &-7.89 &  0.81 & 119.2 &  84.2 & 2.144 & 63.4 & 41.1 &  1.4 & 24.2 &10.4 & 4684.8 &  1.4 & 1.503 &   4.8 & 1.6  &   0.0  & 0.0 &  0.0 &  0.0 & 3.57 & 0.45\\
248 & 2.32 & -2.42 &  50.4 & 223.8 & 2.140 & 73.2 & 61.0 &  1.2 & 23.4 &13.0 & 4684.7 &  1.6 & 2.257 &   8.6 & 2.3  &   0.0  & 0.0 &  0.0 &  0.0 & 3.58 & 0.34\\
256 & 2.79 & -1.61 &  48.4 & 240.0 & 1.867 & 61.4 & 39.4 &  1.4 & 21.8 & 8.9 & 4685.1 &  2.4 & 3.300 &   7.5 & 3.9  &   0.0  & 0.0 &  0.0 &  0.0 & 3.85 & 0.51\\
250 & 1.86 & -2.68 &  49.1 & 214.7 & 1.791 & 79.6 & 68.2 &  1.2 & 25.3 &12.8 & 4684.8 &  1.7 & 1.621 &   7.4 & 2.0  &   0.0  & 0.0 &  0.0 &  0.0 & 3.20 & 0.36\\
234 & 1.39 & -2.42 &  41.9 & 210.0 & 1.667 & 80.9 & 59.2 &  1.3 & 25.6 &15.2 & 4684.7 &  2.0 & 2.191 &  10.9 & 2.9  &   0.0  & 0.0 &  0.0 &  0.0 & 3.19 & 0.29\\
260 & 2.79 & -2.15 &  52.9 & 232.4 & 1.450 & 51.2 & 44.5 &  1.4 & 23.4 & 9.4 & 4684.6 &  2.4 & 1.845 &   6.0 & 2.8  &   0.0  & 0.0 &  0.0 &  0.0 & 3.55 & 0.46\\
238 & 1.39 & -2.95 &  49.1 & 205.2 & 1.290 & 87.0 & 61.5 &  1.3 & 26.8 &16.4 & 4685.0 &  2.5 & 1.594 &   9.4 & 2.7  &   0.0  & 0.0 &  0.0 &  0.0 & 2.93 & 0.27\\
230 & 1.39 & -1.88 &  35.2 & 216.5 & 1.088 & 67.2 & 42.1 &  1.3 & 26.0 &13.6 & 4684.8 &  2.8 & 2.052 &  11.3 & 4.2  &   0.0  & 0.0 &  0.0 &  0.0 & 3.03 & 0.32\\
497 & 3.25 & -0.27 &  49.0 & 265.3 & 1.033 & 49.4 & 28.6 &  1.8 & 23.4 &10.7 &    0.0 &  0.0 & 0.000 &   0.0 & 0.0  &   0.0  & 0.0 &  0.0 &  0.0 & 3.46 & 0.39\\
468 & 1.86 & -1.07 &  32.2 & 240.0 & 0.997 & 49.7 & 26.3 &  1.5 & 24.2 & 9.3 & 4685.1 &  2.6 & 1.294 &   5.4 & 2.9  &   0.0  & 0.0 &  0.0 &  0.0 & 3.31 & 0.47\\
372 & 1.86 & -3.22 &  55.9 & 210.0 & 0.986 & 64.6 & 60.4 &  1.2 & 27.3 &20.6 & 4684.8 &  1.7 & 0.915 &  10.4 & 2.1  &   0.0  & 0.0 &  0.0 &  0.0 & 2.79 & 0.18\\
374 & 2.32 & -3.49 &  63.0 & 213.6 & 0.946 & 54.0 & 50.0 &  1.2 & 25.4 &11.9 & 4684.3 &  0.8 & 0.239 &   2.3 & 0.6  &   0.0  & 0.0 &  0.0 &  0.0 & 3.18 & 0.36\\
404 & 0.31 & -3.76 &  56.5 & 184.7 & 0.904 & 58.2 & 35.1 &  1.3 & 24.9 & 9.5 & 4685.1 &  2.3 & 1.542 &   9.0 & 3.8  &   0.0  & 0.0 &  0.0 &  0.0 & 3.20 & 0.47\\
358 & 2.79 & -3.22 &  64.0 & 220.8 & 0.896 & 47.0 & 47.2 &  1.2 & 25.2 & 9.0 & 4684.8 &  2.3 & 0.850 &   4.8 & 2.1  &   0.0  & 0.0 &  0.0 &  0.0 & 3.21 & 0.49\\
513 & 3.71 & -0.54 &  56.4 & 261.8 & 0.869 & 51.4 & 26.4 &  1.7 & 23.9 & 8.8 &    0.0 &  0.0 & 0.000 &   0.0 & 0.0  &   0.0  & 0.0 &  0.0 &  0.0 & 3.26 & 0.49\\
615 & 0.31 & -8.59 & 129.1 & 182.1 & 0.846 & 55.7 & 78.0 &  1.3 & 28.0 & 8.8 &    0.0 &  0.0 & 0.000 &   0.0 & 0.0  &   0.0  & 0.0 &  0.0 &  0.0 & 2.77 & 0.53\\
236 & 0.93 & -2.68 &  42.7 & 199.1 & 0.843 & 70.3 & 44.7 &  1.4 & 27.2 &15.2 & 4684.8 &  2.5 & 1.393 &  11.8 & 3.7  &   0.0  & 0.0 &  0.0 &  0.0 & 2.80 & 0.28\\
264 & 2.79 & -2.68 &  58.1 & 226.1 & 0.843 & 44.6 & 44.2 &  1.2 & 25.2 &11.0 & 4685.0 &  1.5 & 0.775 &   5.0 & 2.0  &   0.0  & 0.0 &  0.0 &  0.0 & 3.05 & 0.36\\
232 & 0.93 & -2.15 &  35.2 & 203.4 & 0.832 & 70.3 & 29.7 &  1.5 & 24.9 & 9.5 & 4684.8 &  1.1 & 0.584 &   4.4 & 1.6  &4685.0  & 8.2 &  0.7 &  0.1 & 3.08 & 0.48\\
390 & 0.93 & -3.76 &  58.2 & 193.9 & 0.773 & 77.1 & 48.8 &  1.4 & 27.5 &15.1 & 4685.2 &  2.0 & 0.822 &   8.7 & 2.4  &   0.0  & 0.0 &  0.0 &  0.0 & 2.76 & 0.29\\
\hline
418 &  0.31 &  0.31 &   4.7 & 135.0 & 0.175 & 32.0 &  2.3 &  2.5 & 18.9 & 5.4 &    0.0 &  0.0 & 0.000 &   0.0 & 0.0 & 4685.5 & 28.8 & 19.8 &  0.8 & 1.06 & 0.43\\
416 &  0.46 &  0.27 &   8.1 & 120.0 & 0.158 & 31.1 &  1.9 &  2.1 & 22.5 & 6.3 & 4685.1 &  2.5 & 0.413 &   3.4 & 5.8 & 4685.7 & 27.5 & 14.8 &  0.8 & 1.75 & 0.48\\
420 &  0.46 & -0.27 &   8.1 & 240.0 & 0.142 & 45.0 &  2.2 &  2.2 & 18.4 & 4.6 & 4684.5 &  1.7 & 0.326 &   3.3 & 5.1 & 4686.3 & 25.6 & 13.3 &  0.8 & 1.10 & 0.54\\
414 &  0.31 &  0.54 &   8.1 & 150.0 & 0.107 & 33.6 &  2.1 &  2.1 & 22.7 & 5.8 &    0.0 &  0.0 & 0.000 &   0.0 & 0.0 & 4686.6 & 29.8 & 12.6 &  0.6 & 1.44 & 0.51\\
419 & -7.43 &  0.31 & 111.6 &  87.6 & 0.240 & 58.5 & 11.2 &  1.4 & 22.2 &14.6 & 4685.0 &  2.2 & 0.393 &   5.3 & 3.6 & 4685.3 & 73.1 & 11.2 &  0.6 & 1.87 & 0.19\\
422 &  0.31 & -0.54 &   8.1 & 210.0 & 0.098 & 37.2 &  2.6 &  2.3 & 24.7 & 5.4 &    0.0 &  0.0 & 0.000 &   0.0 & 0.0 & 4685.7 & 24.9 & 10.0 &  0.5 & 0.00 & 0.43\\
144 & -0.46 &  0.27 &   8.1 &  60.0 & 0.065 & 29.6 &  2.0 &  2.4 &  3.7 & 1.3 & 4684.5 &  4.2 & 0.363 &   4.5 &12.5 & 4685.4 & 38.7 &  8.3 &  0.5 & 0.00 & 0.00\\
446 &  0.93 &  0.31 &  14.0 & 108.5 & 0.117 & 37.4 &  2.9 &  1.9 & 18.8 & 5.9 & 4685.5 &  1.2 & 0.111 &   2.4 & 2.1 & 4686.7 & 30.0 &  7.2 &  0.5 & 2.50 & 0.51\\
 21 &  0.46 &  0.81 &  14.0 & 150.0 & 0.141 & 47.9 &  3.4 &  2.2 & 17.2 & 6.7 &    0.0 &  0.0 & 0.000 &   0.0 & 0.0 & 4686.5 & 24.5 &  6.9 &  0.5 & 3.38 & 0.49\\
148 & -0.46 & -0.27 &   8.1 & 120.0 & 0.059 & 24.6 &  2.0 &  2.4 & 21.4 & 3.2 & 4683.9 &  1.6 & 0.129 &   2.4 & 4.9 & 4685.9 & 26.6 &  6.5 &  0.4 & 0.00 & 0.65\\
424 &  0.46 & -0.81 &  14.0 & 210.0 & 0.085 & 35.7 &  4.1 &  1.9 & 33.5 & 8.5 & 4684.7 &  2.1 & 0.114 &   2.4 & 3.0 & 4687.1 & 25.6 &  4.4 &  0.3 & 0.00 & 0.35\\
442 &  0.93 &  0.54 &  16.1 & 120.0 & 0.105 & 39.6 &  3.4 &  2.0 & 18.4 & 6.2 & 4685.4 &  1.4 & 0.112 &   2.5 & 2.4 & 4685.8 & 24.1 &  4.4 &  0.4 & 3.36 & 0.55\\
145 & -2.79 &  0.31 &  41.9 &  83.6 & 0.025 & 15.7 &  2.1 &  2.0 &  4.7 & 0.8 &    0.0 &  0.0 & 0.000 &   0.0 & 0.0 & 4685.0 & 28.6 &  3.6 &  0.3 & 0.00 & 0.00\\
152 & -0.46 & -0.81 &  14.0 & 150.0 & 0.034 & 14.1 &  2.2 &  2.3 &  7.9 & 1.7 &    0.0 &  0.0 & 0.000 &   0.0 & 0.0 & 4684.4 & 30.5 &  3.2 &  0.4 & 0.00 & 0.00\\
146 & -0.93 &  0.31 &  14.0 &  71.5 & 0.028 & 12.8 &  1.9 &  2.0 &  3.1 & 0.7 &    0.0 &  0.0 & 0.000 &   0.0 & 0.0 & 4686.4 & 27.2 &  3.0 &  0.3 & 0.00 & 0.00\\
531 & -0.46 &  0.81 &  14.0 &  30.0 & 0.051 & 33.5 &  2.5 &  1.9 & 13.7 & 3.2 & 4684.9 &  1.1 & 0.076 &   2.4 & 3.3 & 4688.7 & 23.1 &  2.9 &  0.3 & 1.85 & 0.66\\
149 & -2.79 & -0.54 &  42.6 & 100.9 & 0.019 & 12.9 &  1.8 &  2.0 &  9.5 & 1.1 & 4688.2 &  2.1 & 0.082 &   0.0 & 9.6 & 4685.8 & 27.8 &  2.6 &  0.3 & 0.00 & 0.00\\
423 & -7.43 & -0.54 & 111.9 &  94.1 & 0.118 & 44.3 & 10.8 &  1.5 & 24.8 & 7.1 & 4684.7 &  1.2 & 0.087 &   2.3 & 1.6 & 4685.6 & 25.8 &  2.5 &  0.4 & 0.48 & 0.39\\
 19 &  0.31 &  1.07 &  16.1 & 163.9 & 0.106 & 40.9 &  4.4 &  2.5 & 18.9 & 4.8 & 4685.2 &  0.7 & 0.071 &   2.2 & 1.5 & 4687.9 & 15.2 &  2.3 &  0.3 & 3.14 & 0.74\\
417 & -7.89 &  0.27 & 118.7 &  88.1 & 0.221 & 71.5 & 12.8 &  1.3 & 23.1 &12.1 &    0.0 &  0.0 & 0.000 &   0.0 & 0.0 & 4684.5 & 10.7 &  2.1 &  0.2 & 1.51 & 0.25\\
 51 & -6.03 &  1.34 &  92.9 &  77.5 & 0.067 & 32.3 & 14.8 &  1.9 & 29.8 & 4.5 & 4684.5 &  1.7 & 0.084 &   2.4 & 2.8 & 4688.6 & 38.4 &  2.0 &  0.3 & 1.75 & 0.96\\
448 &  1.39 & -0.27 &  21.3 & 259.1 & 0.109 & 42.1 &  8.1 &  1.7 & 27.5 & 7.3 &    0.0 &  0.0 & 0.000 &   0.0 & 0.0 & 4689.4 & 25.0 &  1.8 &  0.3 & 1.91 & 0.54\\
452 &  1.39 & -0.81 &  24.2 & 240.0 & 0.335 & 49.2 & 12.2 &  1.7 & 25.6 & 9.4 &    0.0 &  0.0 & 0.000 &   0.0 & 0.0 & 4688.4 & 20.1 &  1.8 &  0.2 & 2.58 & 0.43\\
212 &  0.46 & -1.34 &  21.3 & 199.1 & 0.062 & 28.5 &  7.0 &  1.8 & 37.2 & 6.3 &    0.0 &  0.0 & 0.000 &   0.0 & 0.0 & 4685.4 & 19.5 &  1.5 &  0.2 & 0.04 & 0.57\\
 49 & -5.57 &  1.61 &  87.1 &  73.9 & 0.052 & 28.4 & 11.6 &  2.0 & 31.8 & 3.8 & 4685.4 &  0.9 & 0.056 &   2.6 & 2.4 & 4686.1 & 32.7 &  1.4 &  0.3 & 1.63 & 1.18\\
\hline
\end{tabular}
\end{tiny}
\begin{tablenotes}
\item 
Row description: \\
Data for 50 fibres, ordered according to the decreasing contribution in \hb\
(column 6) in the first 25 rows, and ordered according to the decreasing number of WR stars (column 19)
in the next 25 lines. Table containing data for all 567 fibres (excluding the sky fibres) is available in the
electronic version.\\
Column description: \\
1 - Fibre number. \\
2--3 - X and Y shifts in arcsecs with respect to (wrt) the core of SSC-A, which is the closest to fibre~\#418. 
     Objects to the east and south of SSC-A have negative shifts. \\
4 - distance from SSC-A in parsec using a scale of 15~pc\,arcsec$^{-1}$. \\
5 - PA in degrees of the fibre position wrt SSC-A. \\
6 - 100 $\times$ the extinction-corrected \hb\ flux in the fibre wrt total \hb\ flux (4.444$\times10^{-12}$~\ergcms).\\
7 - Signal-to-noise ratio (SNR) of the \hb\ line (see equation 1).\\
8 - Emission Equivalent Width (EW) in \AA\ of the \hb\ emission line. \\
9 - FWHM in \AA\ of the fitted Gaussian to the emission line of \hb. \\
10 - 100$\times\frac{f_{H\gamma}}{f_{H\beta}}$ ratio.\\
11 - SNR of the \hg\ emission line.\\
12 - Central wavelength in \AA\ of the fitted \heiiwr\ narrow (nebular) line.\\
13 - FWHM in \AA\ of the fitted \heiiwr\ narrow (nebular) line.\\
14 - 100 $\times$ the extinction-corrected \heiiwr\ nebular line flux in the fibre wrt total flux in the same line 
    (9.13$\times10^{-14}$~\ergcms).\\
15 - SNR of the fitted \heiiwr\ narrow (nebular) line. \\
16 - 100$\times\frac{f_{He{\sc ii}\lambda4686(nebular)}}{f_{H\beta}}$ ratio.\\
17 - Central wavelength in \AA\ of the fitted \heiiwr\ broad (WR) line.\\
18 - FWHM in \AA\ of the fitted \heiiwr\ broad (WR) line.\\
19 - Luminosity of the \heiiwr\ broad line expressed in units of 1.22$\times10^{36}$~\ergs\ $\equiv N_{\rm WR}$=equivalent number of WNL stars.\\
20 - Error in $N_{\rm WR}$, determined by propagating errors in the \heiiwr\ broad line flux.\\
21 - \Av\ determined from \hg\ and \hb\ lines when both the lines have $SNR>3$ (absorption correction 1~\AA\ in EWs.)\\
22 - Error on \Av, determined by propagating errors in the \hb\ and \hg\ line fluxes.
\end{tablenotes}
\end{threeparttable}
\end{table*}

\end{document}